\documentclass{aastex}
\usepackage{emulateapj5}
\usepackage{apjfonts}
\usepackage{epsf}
\usepackage{onecolfloat}

\begin{document}
\twocolumn[
\title{Density profiles of  $\Lambda$CDM clusters}
\author{Argyro Tasitsiomi\altaffilmark{1}, Andrey V. Kravtsov\altaffilmark{1}, 
Stefan Gottl\"ober\altaffilmark{2}, Anatoly A. Klypin\altaffilmark{3}}



\begin{abstract}
  We analyze the mass accretion histories (MAHs) and density profiles
  of cluster-size halos with virial masses of $0.6-2.5\times
  10^{14}h^{-1}\rm\ M_{\odot}$ in a flat $\Lambda$CDM cosmology. We
  find that most MAHs have a similar shape: an early, merger-dominated
  mass increase followed by a more gradual, accretion-dominated
  growth.  For some clusters the intense merger activity and rapid
  mass growth continue until the present-day epoch.  In agreement with
  previous studies, we find that the concentration of the density
  distribution is tightly correlated with the halo's MAH and with its
  formation redshift.  During the period of fast mass growth the
  concentration remains approximately constant and low $c_{\rm
    v}\approx 3-4$, while during the slow accretion stages the
  concentration increases with decreasing redshift as $c_{\rm
    v}\propto (1+z)^{-1}$.  We consider fits of three widely discussed
  analytic density profiles to the simulated clusters focusing on the
  most relaxed inner regions.  We find that there is no unique best
  fit analytic profile for all the systems. At the same time, if a
  cluster is best fit by a particular analytic profile at $z=0$, the
  same is usually true at earlier epochs out to $z\sim 1-2$. The local
  logarithmic slope of the density profiles at $3 \%$ of the virial
  radius ranges from $-1.2$ to $-2.0$, a remarkable diversity for the
  relatively narrow mass range of our cluster sample.  Interestingly,
  for all the studied clusters the logarithmic slope becomes shallower
  with decreasing radius without reaching an asymptotic value down to
  the smallest resolved scale ($\lesssim 1\%$ of the virial radius).
  We do not find a clear correlation of the inner slope with the
  formation redshift or the shape of the halo's MAH. We do find,
  however, that during the period of rapid mass growth the density
  profiles can be well described by a single power law $\rho(r)\propto
  r^{-\gamma}$ with $\gamma\sim 1.5-2$.  The relatively shallow power
  law slopes result in low concentrations at these stages of
  evolution, as the scale radius where the density profiles reaches
  the slope of $-2$ is at large radii.  This indicates that the inner
  power law like density distribution of halos is built up during the
  periods of rapid mass accretion and active merging, while outer steeper
profile is formed when the mass accretion slows down. To check the
  convergence and robustness of our conclusions, we resimulate one of
  our clusters using eight times more particles and twice better force
  resolution.  We find good agreement between the two simulations in
  all of the results discussed in our study.
 
\end{abstract}


]
\altaffiltext{1}{Department of Astronomy and Astrophysics,
       Center for Cosmological Physics,
       The University of Chicago, Chicago, IL 60637;
       {\tt iro, andrey@oddjob.uchicago.edu}}
\altaffiltext{2}{Astrophysikalisches Institut Potsdam
An der Sternwarte 16, 14482 Potsdam, Germany; {\tt sgottloeber@aip.de}}
\altaffiltext{3}{Department of Astronomy, New Mexico State University,
P.O. Box 30001, Las Cruces, NM, 88003; {\tt aklypin@nmsu.edu}}

\section{Introduction}
\label{sec:intro}

During the last decade, there has been an increasingly growing
interest in testing the predictions of variants of the cold dark
matter models (CDM) on small scales. The interest was spurred by
indications that the density distribution in the inner regions of dark
matter halos predicted by CDM is at odds with the observed galactic
rotation curves \citep{flores_primack94,moore94}. This discrepancy is
yet to be convincingly resolved and is still a subject of active
debate
\citep[e.g.,][]{cote_etal00,vandenbosch_swaters01,
blais_ouellette_etal01,deblok_etal01,deblok_etal03,swaters_etal03}.
In addition, the CDM models face other apparent discrepancies with
observations on galactic scales such as the amount of substructure in
galactic halos \citep{klypin_etal99,moore_etal99}, the incorrect
normalization of the Tully-Fisher relation, the angular momentum of
disk galaxies \citep{navarro_steinmetz97,navarro_steinmetz00}, the
ellipticity of dark matter halos \citep{ibata_etal01}, and others.

In the past several years, the density distribution in the cores of
galaxy clusters has also become a subject of a related debate. 
CDM models predict cuspy density profiles without flat cores
\citep{frenk_etal85,quinn_etal86,dubinski_carlberg91}. 
\citet[NFW]{nfw_96,nfw_97} argued that the CDM
halo profiles can be described by the following simple formula in all
cosmologies and at all epochs,
\begin{equation}
\label{eq:nfw}
\rho(r)=\frac{\rho_{0}}{(r/r_{s})(1+r/r_{s})^{2}}.
\end{equation}
This analytic formula describes the density profile of a halo using
two parameters: a characteristic density, $\rho_{0}$, and a
scale radius, $r_{s}$. These parameters are determined by the halo
virial mass, $M_{v}$, and concentration index, $c \equiv r_{v}
/r_{s}$, where $r_{v}$ is the virial radius of the halo. In addition,
NFW argued that there is a tight correlation
between $c$ and $M_{v}$, which means that the halo profiles of different
mass objects form a one parameter family.  

\citet{moore_etal98} \citep[see also][]{ghigna_etal00} carried out a
convergence study of the dark matter profiles and concluded that high
mass resolution is required to resolve the inner density distribution
robustly. They advocated the analytic density profile of the form
$\rho(r) \propto (r/r_{s})^{-1.5}[1+(r/r_{s})^{1.5}]^{-1}$, as a
better description of the density distribution of their simulated
halos.  This profile behaves similarly to the NFW profile at large
radii ($\propto r^{-3}$), but is steeper at small radii ($\propto
r^{-1.5}$).
\citet{fukushige_makino97,fukushige_makino01,fukushige_makino03}
reached similar conclusions using a set of independent simulations.
\citet{jing_suto00} presented a systematic study of the density
profiles of halos with masses in the range $2 \times 10^{12}-5\times
10^{14} \ h^{-1} \ M_{\odot}$.  They found that the inner slope at a
radius of 1\% of the virial radius is shallower ($\approx -1.1$) for
cluster mass halos than for galactic halos ($\approx -1.5$).  
Recently, \citet{hayashi_etal03, navarro_etal03} found that often 
the logarithmic slope of the density distribution
at the convergence radius 
is steeper than -1 as expected from the NFW profile, but
significantly shallower than the -1.5  inner slope 
found by \citet{moore_etal98}.
Several
other studies \citep{kravtsov_etal97, kravtsov_etal98,
  avila_reese_etal99, jing00, bullock_etal01,
  klypin_etal01,fukushige_etal03} found a significant scatter in both
the shape of the density profiles and halo concentrations, likely
related to the details of the mass accretion histories of individual
objects \citep{wechsler_etal02,zhao_etal03a}. \citet{mucket_hoeft03}
and \citet{hoeft_etal03} study the radial dependence of the
gravitational potential and the velocity dispersion and come to the
conclusion that there does not exist a slope asymptote of the density
profile over a wide range but the slope increases with decreasing
radius and reaches the value -0.58 as $r \rightarrow 0$.

Observational constraints on the inner slope of the dark matter
density distribution in galactic halos are difficult because the
distribution is affected by the cooling and dynamics of baryons. The
dark matter profiles in clusters, on the other hand, should be less
affected by cooling as a much smaller fraction of cluster baryons is
observed to be in the cold condensed phase.  Observational studies
of the mass distribution in clusters using weak lensing and
hydrostatic analysis of the X-ray emitting gas show that the overall mass
distribution is in general agreement with CDM predictions
\citep{allen_98,clowe_etal00,willick_padmanabhan00,
clowe_schneider01,sheldon_etal01,arabadjis_etal02, athreya_etal02,
bautz_arabadjis03}. 

Strong lensing studies can probe the mass distribution in the inner
region of clusters and thus test the ``cuspiness'' of cluster halos.
However, the results of strong lensing analyses have so far been
contradictory, even in the case when the same system was studied.
\citet{tyson_etal98}, for example, argue that the density profile of
cluster CL0024$+$1654 has a constant density core, while
\citet{broadhurst_etal00} find that the mass distribution in this
cluster is cuspy. \citet{czoske_etal02} argue that CL0024$+$1654 is
undergoing a major merger and its density profile may not be
representative.  \citet{sand_etal02} find that the inner slope of the
density profile in cluster MS$2137-23$ is flatter than expected in CDM
models, a conclusion they recently confirmed for six more clusters
\citep{sand_etal03}. \citet{gavazzi_etal03}, reanalyzing the same
observations, argue that if the fifth demagnified image near the
center of the lensing potential is not taken into account then the
inner slope may be consistent with CDM predictions.

Given the disagreement among the different analytical fits proposed
for the density profiles of dark matter halos found in simulations and
a possible discrepancy with strong lensing observations, it is
interesting to conduct a systematic study of the density profiles of
clusters in the concordance $\Lambda$CDM model. The study of cluster
mass halos is also interesting because the typical concentrations of
their matter distribution are lower than those of galactic halos. Thus, if
an asymptotic inner slope, suggested by the analytic profiles, does
exist it should be reached at a larger fraction of the virial radius
in cluster halos and should be easier to detect.

The paper is organized as follows.  In the following two sections, we
describe the numerical simulations and halo finding algorithm 
used in our analysis. In
$\S$~\ref{sec:mah} we discuss the mass accretion histories of the
analyzed clusters. In $\S$~\ref{sec:density_prof} we present the 
convergence test, discuss the fitting procedure, and 
our results on the shapes and inner slopes of the density profiles. 
We summarize
our results and conclusions in $\S$~\ref{sec:conclusions}.

\section{Numerical Simulations}
\label{sec:sims}
We use the Adaptive Refinement Tree code \citep{kravtsov_etal97} to
follow the evolution of cluster-size halos in the flat $\Lambda$CDM
cosmology: $(\Omega_{\rm m},\Omega_{\Lambda},h,\sigma_8)
$$=($$0.3$,$0.7$,$0.7$,$0.9)$. We use the initial spectrum in the
Holtzman approximation with $\Omega_b=0.03$ \citep[see][]{klypin_holtzman97}.
The code starts with a uniform $256^3$ grid covering the entire
computational box. This grid defines the lowest (zeroth) level of
resolution.  Higher force resolution is achieved in the regions
corresponding to collapsing structures by recursive adaptive
refinement of all such regions. Each cell can be refined or de-refined
individually. The cells are refined if the particle mass contained
within them exceeds a certain specified threshold value. The code thus
refines to follow the collapsing objects in a quasi-lagrangian
fashion.

The cluster halos were simulated in a box of $80h^{-1}$ Mpc.  A low
resolution simulation was run first. A dozen cluster halos were
identified and multiple mass resolution technique was used to set up
initial conditions \citep{klypin_etal01}.  Namely, a lagrangian region
corresponding to a sphere of radius equal to two virial radii around
each halo was re-sampled with the highest resolution particles of mass
$m_{\rm p}=3.16\times 10^8h^{-1}{\rm M_{\odot}}$, corresponding to an
effective number of $512^3$ particles in the box, at the initial redshift 
of the simulation ($z_{\rm i}=50$). The high mass resolution region was
surrounded by layers of particles of increasing mass with a total of
three particle species. Only regions containing highest resolution
particles were adaptively refined and the threshold for refinement was
set to correspond to a mass of $4$ highest resolution particles per
cell.  Each cluster halo is resolved with $\sim 10^6$ particles within
its virial radius at $z=0$. The size of the highest refinement level
cell was $1.2h^{-1}$~kpc.  In addition, one of the clusters was
re-simulated with eight times more particles ($m_{\rm p}=3.95\times
10^7h^{-1}{\rm M_{\odot}}$) to study the convergence of the density
profiles. In this simulation, the smallest cell size reached was
$0.6h^{-1}$~comoving kpc. 

The time steps were chosen so that no particle moves by more than a
fraction of the parent cell size in a single step. This criterion was
motivated by the convergence studies presented by
\citet{klypin_etal01}.  For the analyzed simulations, the number of
steps at the highest refinement level was $\approx 250000$ or $\Delta
t\approx 2-3\times 10^4$~yrs and a factor of 2 larger for each lower
refinement level. For the high-resolution resimulation of one of the
cluster used for convergence check, the number of steps at the highest
refinement level was $\approx 500000$.  We analyze the cluster
profiles and their mass accretion histories using 19 outputs from
$z=10$ to $z=0$, with a typical time interval between outputs of $\sim
0.7$~Gyr.

\begin{table}[tb]
\tablenum{1}
\caption{Simulated Halo parameters \label{tab:clp}}
\begin{center}
\small
\begin{tabular}{lccc}
\tableline\tableline\\
\multicolumn{1}{l}{Halo } & 
\multicolumn{1}{c}{$M_{180}$}&
\multicolumn{1}{c}{$r_{180}$} &
\multicolumn{1}{c}{$V_{\rm max}$} 
\\
& ($h^{-1} \ M_{\odot}$)  &    ($h^{-1} \rm{Mpc}$)     & ($\rm{km} \ \rm{s}^{-1}$)   \\
\\
\tableline
\\
CL1 &    $2.5 \times 10^{14}$ &   $1.58$    &   $973$ \\
CL2 &    $2.4 \times 10^{14}$ &   $1.56$    &   $1011$ \\ 
CL3 &    $2.3 \times 10^{14}$ &   $1.55$    &   $904$\\
CL4 &    $1.3 \times 10^{14}$ &   $1.29$    &   $826$  \\ 
CL5 &   $1.3 \times 10^{14}$ &   $1.27$    &   $798$ \\
CL6 &   $1.2 \times 10^{14}$ &   $1.25$    &   $785$ \\
CL7 &   $1.2 \times 10^{14}$ &   $1.23$    &   $587$ \\
CL8 &   $1.2 \times 10^{14}$ &   $1.23$    & $695$ \\ 
CL9 &    $9.7 \times 10^{13}$ &   $1.16$    &   $630$  \\
CL10 &   $8.6 \times 10^{13}$ &   $1.11$    &   $597$  \\ 
CL11 &   $8.1 \times 10^{13}$ &   $1.09$    &   $670$ \\
CL12 &   $8.1 \times 10^{13}$ &   $1.09$    &   $758$ \\  
CL13 &   $7.3 \times 10^{13}$ &   $1.05$    &   $607$ \\
CL14 & $5.8 \times 10^{13}$ & $0.98$ & $603$\\ 
\\             
\tableline
\end{tabular}
\end{center}
\end{table}

\section{Halo Identification}
\label{sec:haloid}

To identify cluster halos we use a variant of the Bound Density Maxima
(BDM) halo finding algorithm. The main idea of the BDM algorithm is to
find positions of local maxima in the density field smoothed at a
certain scale and to apply physically motivated criteria to test
whether the identified site corresponds to a gravitationally bound
halo. The detailed description of the algorithm is given in \citet{klypin_holtzman97} and \citet{klypin_etal99}.  

We start by calculating the local overdensity at each particle
position using the SPH smoothing kernel\footnote{To calculate the
density we use the publicly available code {\tt smooth:
http://www-hpcc.astro.washington.edu/tools/tools.html}} of 24
particles.  We then sort particles according to their overdensity and
use all particles with $\delta \geq \delta_{\rm min}=5000$ as
potential halo centers.  Starting with the highest overdensity
particle, we surround each potential center by a sphere of radius
$r_{\rm find}=50h^{-1}\ \rm kpc$ and exclude all particles within this
sphere from further center search.  After all the potential centers are
identified, we analyze the density distribution and velocities of the
surrounding particles to test whether the center corresponds to a
gravitationally bound clump \citep{klypin_etal99}. We then construct
profiles using only bound particles and use them to calculate the
properties of halos such as the maximum circular velocity $V_{\rm
  max}$, the mass $M$, etc. In this study, we consider only
isolated cluster-size halos. We should note that for isolated halos 
the BDM algorithm works very similarly to the commonly used spherical
overdensity (SO) algorithm. 

The virial radius is a convenient measure of the halo size. We define
the virial radius as the radius within which the density is equal to
180 times the {\it average} density of the universe at a given epoch.
The separation between halos is sometimes smaller than the sum of
their virial radii. In such cases, the definition of the outer
boundary of a halo and its mass are somewhat ambiguous.  To this end,
in addition to the virial radius, we estimate the {\rm truncation
  radius}, $r_{\rm t}$, at which the logarithmic slope of the density
profile constructed from the bound particles becomes larger than
$-0.5$ as we do not expect the density profile of the CDM halos to be
flatter than this slope. In general we consider the halo radius to be
$r_{\rm h}=\min(r_{\rm 180},r_{\rm t})$.

In our analysis we use only clusters with masses \linebreak $>5 \times
10^{13} h^{-1} \ M_{\odot}$.  In most cases two or more clusters were
identified with this mass threshold in each run.  To distinguish
between the isolated cluster halos and massive subhalos we use
additional information, such as the virial-to-tidal radius ratio, the
maximum circular velocity, and the number of gravitationally bound
particles within $r_{\rm h}$.  We consider halos to be isolated if
their separation is larger than one third of the sum of their virial
radii. We list the present-day properties of the cluster halos
included in our sample in Table~\ref{tab:clp}. The masses and radii
correspond to the cumulative overdensity of 180 times the mean density
of the Universe. In this list, clusters 4/12 and 10/13 are close
pairs, clusters 6/11/14 and 7/8/9 are triplets, while clusters 1, 2,
3, and 5 are well-isolated systems. The clusters thus sample a variety
of environments.

We stress that the clusters in the analyzed sample were selected
randomly -- no specific criterion of relaxation or substructure was
used. As the Figure~1 below shows clusters in our sample span a wide range of
mass accretion histories and formation redshifts.

\section{Mass accretion histories}
\label{sec:mah}

In the hierarchical structure formation scenario 
halos are assembled via a
continuous process of merging and accretion. Details of the mass
accretion history (MAH) may affect the shape of the halo and its
density distribution \citep{nfw_97,bullock_etal01,wechsler_etal02,zhao_etal03a}.
It is therefore interesting to study the accretion history of the
halos in conjunction with the study of their density profiles.  In
this section we study the details of the assembly of the simulated
cluster halos by following the most massive progenitor from $z=10$
up to the present. We will discuss connections between the halo 
density profile and its MAH in \S~\ref{results} and \ref{sec:conclusions}.

\subsection{Constructing MAHs}
\label{constructing_mah}

For each $z=0$ cluster halo we identify the most massive progenitor
using the halo catalogs of the previous time output.  In what follows,
if halo 1 is the most massive {\em progenitor\/} of halo 2, then halo
2 will be referred to as the {\em offspring\/} of halo 1.  To identify
the most massive progenitor of a halo we first identify all of its
progenitors in the halo catalog. We then eliminate  from
the set of potential progenitors, objects that
are  significantly tidally stripped (i.e., their virial-to-tidal 
radius ratio is greater than 3.5).  We use the following criteria to
identify the most massive progenitor among the remaining candidates.
1. We eliminate candidate progenitors with masses less than 20\% of
the offspring mass. 2. Using the peculiar velocity of the offspring,
we find the approximate location of the progenitor in the previous
output and eliminate the candidates outside the sphere with radius
$r=10 \times v_p \Delta t$, where $v_p$ is the offspring peculiar
velocity and $\Delta t$ the time elapsed between two successive
outputs.  3. We require that candidate progenitor and offspring halos
have a certain fraction of common particles.  In what follows, $f_1$
($f_2$) denotes the ratio of the number of particles that offspring
and progenitor have in common to the number of particles in the
offspring (progenitor).  

\begin{figure*}[tp]
\centerline{\epsfxsize=\textwidth\epsffile{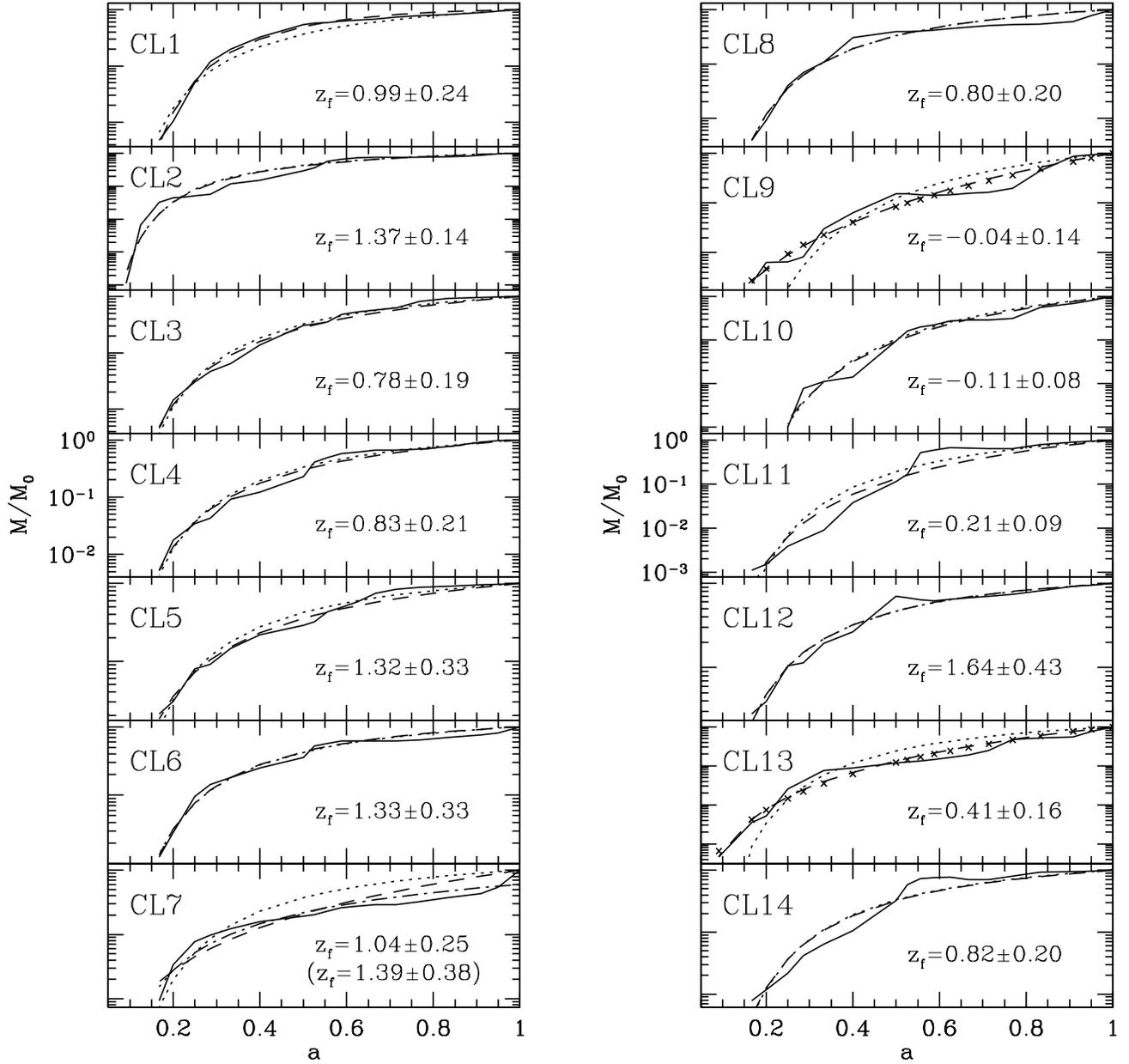}}
\caption{Mass accretion histories of the cluster halos ({\it solid lines}). 
  Also shown are the analytic fits of Eq.~(\ref{eq:our_function}),
  $M(\tilde{a})/M_0= \tilde{a}^p\exp[-\alpha(\tilde{a}-1)]$ (where
  $\tilde{a}\equiv a/a_0$, and $a=(1+z)^{-1}$). The {\it dashed lines}
  show fits with both $\alpha$ and $p$ varied, while the {\it dotted
    lines} show the fits with the parameter $p$ fixed to zero. The
  formation redshift $z_{f}$, given by Eq.~(\ref{eq:form_z}) in terms
  of $\alpha$, is shown in the legend of each panel.  The best fit
  $\alpha$ obtained from the two fits is nearly identical in all
  cases, except for CL9 and CL13.  For CL9 and 13 we also plot the fit
  obtained using Eq.~(\ref{eq:vandenbosch}) ({\it crosses}).  Finally,
  to show the effect of a recent major merger, for CL7 we plot the
  fit assuming an epoch of observation $a_{0}=0.95$ rather than
  $a_{0}=1$, used for all the other fits ({\it dash-dotted line}). The
  $z_f$ value obtained in this case is given in the parentheses.
\label{fig:growth_curves}}
\vspace{-0.5cm}
\end{figure*}

The candidate with the largest number of common particles with the
offspring and with $f_1\ge 0.5$ is then chosen to be the most massive
progenitor.  At the same time the condition $f_2 \ge 0.5$ is also
checked and found to be satisfied. If all the progenitors have $f_1<
0.5$, as is often the case during major mergers, for the progenitor
with the largest $f_1$ we also require that more than $95 \%$ of the
particles within a comoving radius of $10 \ h^{-1} \ \rm{kpc}$ from
the most bound particle of the progenitor are also found in the
offspring. Starting from $z=0$, we repeat the identification of the
most massive progenitor for all the 19 simulation output epochs to $z
\simeq 10$, or until the progenitor can no longer be identified. 
The mass accretion histories
constructed in this way for each of the analyzed clusters are shown in
Figure~\ref{fig:growth_curves} (solid lines).

Most of the MAHs have a qualitatively similar shape: a rapid increase
in mass during the early epochs and a relatively slow increase at the
later stages of evolution.  Despite the similarities, the details of
MAHs differ significantly from object to object.  CL9, CL10, and CL13
have not yet reached the second, accretion dominated stage of their
evolution.  Their mass is accumulated via intense merger activity up
to the present epoch. CL7 and CL8 appear to have reached the slow
accretion phase, but experienced a late major merger.  The masses of
CL11 and 14 at early epochs increase almost linearly with the
expansion factor ($\log{(M/M_{0})} \propto a$), and reach a
$M\approx\rm const$ plateau at later epochs.

\subsection{Major mergers}

In addition to the overall shape of the halo MAH, it is useful to have
some more specific information on the major mergers experienced by the
clusters. We will use the term {\it major merger} to describe all
events that result in a more than $30\%$ increase in the mass of the
main progenitor between the two output epochs.  
As we
mentioned above, the average time elapsed between two successive
outputs of the simulation is $\approx 0.7$ Gyr, which is close to the
crossing time of $\approx 1$~Gyr for a wide range of halo masses. The
crossing time is a lower limit for the merger time scale, which means
that the spacing of our outputs is appropriate for merger identification
\citep[see also tests in][]{gottloeber_etal01}.
We tabulate the redshift of the last major merger, $z_{LMM}$,
as well as the corresponding fractional mass change, $\Delta M/M$, for
all the clusters in columns 2 and 3 of Table~\ref{tab:z},
respectively. One should keep in mind that these numbers are only indicative,
since defining a major merger, e.g., as a $20 \%$ mass increase, would render $z_{LMM}$ for CL4 
equal to $\simeq 0.15$. 

\begin{table*}[tb]
\tablenum{2}
\label{tab:z}
\caption{Halo parameters from mass accretion histories and density fits}
\begin{center}
\small
\begin{tabular}{lccrrcrc}
\tableline\tableline\\
\multicolumn{1}{c}{Halo } & 
\multicolumn{1}{c}{$z_{LMM}$}&
\multicolumn{1}{c}{$\Delta M /M$} &
\multicolumn{1}{c}{$z_{1/2}$} &
\multicolumn{1}{c}{$z_{f}$} &
\multicolumn{1}{c}{Best fit} &
\multicolumn{1}{c}{$c_{-2}$}  &
\multicolumn{1}{c}{slope}  
\\
\multicolumn{1}{c}{(1)} & 
\multicolumn{1}{c}{(2)}&
\multicolumn{1}{c}{(3)} &
\multicolumn{1}{c}{(4)} &
\multicolumn{1}{c}{(5)} &
\multicolumn{1}{c}{(6)} &
\multicolumn{1}{c}{(7)} & 
\multicolumn{1}{c}{(8)}  
\\
\tableline
\\
CL1 &    $1.24$   &   $0.65$    &   $1.09$ & $0.99 \pm 0.24$  & M     & $9.7$   & $-1.23 \pm 0.19$ \\
CL2 &    $0.85$   &   $0.56$    &   $0.83$ & $1.37 \pm 0.14$  & JS     & $9.6$  & $-1.67 \pm 0.15$\\ 
CL3 &    $0.75$   &   $0.36$    &   $0.67$ & $0.78 \pm 0.19$  & NFW     & $10.7$   & $-1.35 \pm 0.20$ \\
CL4 &    $0.95$   &   $0.77$    &   $0.78$ & $0.83 \pm 0.21$  & JS     & $5.4$   & $ -1.89 \pm 0.20$ \\ 
CL5 &    $0.85$   &   $0.34$    &   $0.69$ & $1.32 \pm 0.33$  & M     & $12.2$   & $-1.30 \pm 0.20$ \\
CL6 &    $0.95$   &   $0.45$    &   $0.91$ & $1.33 \pm 0.33$  & JS     & $14.7$  & $-1.83 \pm 0.19$ \\
CL7 &    $0.03$   &   $0.82$    &   $0.07$ & $1.04 \pm 0.25$  & JS     & $13.6$ & $-2.01 \pm 0.22$\\
CL8 &    $1.74$   &   $1.83$    &   $0.41$ & $0.80 \pm 0.20$  & M     & $11.5$  & $-1.25 \pm 0.22$  \\
CL9 &    $0.15$   &   $0.95$    &   $0.19$ & $-0.04 \pm 0.14$ & M     & $3.5$   & $-1.68 \pm 0.22$\\
CL10 &   $0.25$   &   $0.78$    &   $0.22$ & $-0.11 \pm 0.08$ & M     & $2.3$   & $-1.78 \pm 0.28$ \\ 
CL11 &   $0.85$   &   $2.11$    &   $0.80$ & $0.21 \pm 0.09$  & NFW     & $8.4$  & $-1.38 \pm 0.31$  \\
CL12 &   $1.24$   &   $1.67$    &   $1.23$ & $1.64 \pm 0.43$  & M     & $12.6$  & $-1.50 \pm 0.24$ \\  
CL13 &   $0.03$   &   $0.33$    &   $0.25$ & $0.41 \pm 0.16$  & JS     & $4.3$   & $-1.36 \pm 0.42$\\
CL14 &   $0.95$   &   $0.81$    &   $0.93$ & $0.82 \pm 0.20$  & JS     & $10.8$  & $-1.42 \pm 0.30$\\
\\             
\tableline
\end{tabular}
\tablecomments{(2): redshift of last major merger, (3): fractional mass change during last major merger,
(4): redshift where half of the cluster's current mass has been accreted, (5): formation redshift defined
from MAHs, (6): best fit among the \citet{nfw_96, nfw_97}, the \citet{moore_etal98}, and
the \citet {jing_suto00} profiles (7): concentration index, 
(8): logarithmic slope as obtained by averaging the local logarithmic slope
between the smallest resolved radius and $3\%$ of the virial radius}
\end{center}
\end{table*}

\subsection{Formation redshift and MAH shape}

To characterize evolution of the halos, one can introduce the halo
formation epoch (or redshift).  Usually, the formation epoch is defined
as the time when the mass in the most massive progenitor(s) is equal to
some fraction of the halo's final mass, $M_{0}$ 
\citep[e.g., see][]{lacey_cole93,nfw_97}.  Taking this fraction to be equal to
$1/2$ we calculate the formation redshift, $z_{1/2}$, which we
tabulate in column 4 of Table~\ref{tab:z}. In order to find $z_{1/2}$
we use linear interpolation between the successive outputs that bracket
$M/M_{0}=1/2$.  It is interesting to note that $z_{1/2}$ is typically
smaller than $z_{LMM}$.

As pointed out by \citet[hereafter W02]{wechsler_etal02}, defining
 the formation redshift as the redshift where the ratio $M/M_{0}$
takes a specific value gives a formation redshift that 
 depends on the time of observation of the halo. In addition, the
definition uses the MAH of the halo at two epochs only (the formation and
present epochs) and therefore makes it sensitive to the local jumps in
the MAH and less sensitive to the overall MAH shape. The case of CL7 
may serve as an illustration. This object had
entered its quiescent stage of evolution relatively early.
Nevertheless, the low value of its formation redshift, $z_{1/2}$, is
determined largely by the single late major merger.  In view of these
considerations, W02 argued that 
the mass accretion histories can be better characterized by a formation
redshift that is derived from a functional fit to the entire MAH. 
Namely, they propose to fit the MAHs of halos by a simple exponential:
\begin{equation}
\label{eq:w02mah}
\tilde{M}(\tilde{a})=\exp \left [\alpha \left (1-1/\tilde{a}\right ) \right ];\ \ \tilde{a}\equiv a/a_0;\ \ a=(1+z)^{-1},
\end{equation}
where $\tilde{M}\equiv M/M_0$, and $M_0$ and $a_0$ are the virial mass of the halo and expansion 
factor at the epoch of observation, respectively. 
Using the fit, one can define the formation
epoch independent of the epoch of observation as the 
redshift corresponding to a fixed value of 
$d \log{M} / d \log{a}=S$. The value of $S$ is arbitrary,
and we follow W02 and choose $S=2$ since this is the value
required to match the concentration index-collapse 
redshift relation found by \citet{bullock_etal01}. 
The formation redshift, $z_{f}$, can then be defined by the relation
\begin{equation}
\label{eq:form_z}
z_{f}= \frac{2}{\alpha}(1+z_{0}) -1 .
\end{equation}
The formation redshift in this definition is independent of the epoch 
of observation;
the factor $(1+z_{0})$ appears only in order to 
comply with the convention of $z=0$ corresponding to the present epoch. 

Although the simplicity of the above expression
is attractive, we find that in some cases it provides a rather poor
fit to the individual MAHs. 
We generalize the fitting formula by the following two-parameter function
\begin{equation}
\label{eq:our_function}
\tilde{M}(\tilde{a})=\tilde{a}^{p}\exp \left [\tilde{\alpha} \left (1-1/\tilde{a}\right) \right ],
\end{equation}
The function in Eq.~(\ref{eq:w02mah}) 
is a special case of Eq.~(\ref{eq:our_function}) for $p=0$. 
The fits in which both $\alpha$ and $p$ were varied and the fits with
fixed $p=0$ are shown in Figure~\ref{fig:growth_curves}. These fits were
obtained by $\chi^2$ minimization, even though the robustness of their relative 
quality with respect to the choice 
of merit function was tested. For CL9 and
13 the fits with $p=0$ are rather poor.  For the two-parameter fits to
the MAHs of these clusters the value of $\tilde{\alpha}$ is close to
zero, which means that the MAHs are better described by a power law
in $\tilde{a}$, rather than by an exponential in $1/\tilde{a}$.  
Detailed study of the MAH shapes
clearly requires a larger sample of halos. Analyzing the 
merger histories in terms of number of major mergers
indicates that the power law behavior 
may be related to the high frequency of major mergers 
up to the present epoch. We note that the {\em galaxy-size} halos studied
by W02 formed earlier on average than our halos, and thus
only a small fraction ($<5 \%$) of their halos were similar to our
CL9 and 13. 

Clearly, a smooth fitting function for the MAHs cannot capture all the
features of the actual evolution, such as minor and major merger
events. These events however do influence the values of the 
best fit parameters. In the case of CL7, one can
see the effect of a very recent major merger. As can be seen in Figure
\ref{fig:growth_curves}, if we make the fit at 
$a_{0}=0.95$ instead of $a_0=1$ (i.e., with the observation epoch 
prior to the merger), we get a better overall fit. 

\citet{vandenbosch_02} found that the {\it average} mass accretion history
of halos generated using the extended Press-Schechter (EPS) formalism 
is described well by a different two-parameter
function
\begin{equation}
\label{eq:vandenbosch}
\log{\langle\tilde{M} \rangle}=-0.301 (-1)^{\nu}
\left [\frac{\log{(a)}}{\log{(1+\tilde{z}_f)}}
\right ]^{\nu},
\end{equation}
with $\tilde{z}_f, \nu$ the parameters to be determined.  By
definition, $\tilde{z}_f$ is the redshift which corresponds to
$\langle M/M_{0} \rangle =1/2$.  Typical best fit values for $\nu$ are
in the range of $1.4-2.3$. As before, clusters CL9 and 13 are
exceptions with $\nu \approx 1$ (i.e.,  their MAH is power-law $M
\approx M_{0} a^{\tilde{p}}$). We find that Eq.~(\ref{eq:vandenbosch})
gives fits equally good to those obtained with the two-parameter
function of Eq.~(\ref{eq:our_function}) in all cases. This function
however is not convenient when used to calculate the formation redshift via the
logarithmic derivative of the mass with respect to the scale factor.
More specifically, in  Eq.~(\ref{eq:vandenbosch}) (and its derivative) 
 $\tilde{z_{f}}$ is by construction  positive, and thus for a general 
 $\nu$,  $a$ has to be $\le 1$.  This will not be the case for objects 
 whose mass accretion rate  reaches  $S=2$ in the future (negative formation redshift), 
 and thus Eq.~(\ref{eq:vandenbosch}) cannot be used in 
 these cases to obtain a formation redshift.
We
choose to use Eq.~(\ref{eq:our_function}), as it is a simple extension of the function
used by W02 and the definition of $z_f$ via the mass logarithmic derivative is 
guaranteed.

The formation redshifts estimated using the best fit parameters of
Eq.~(\ref{eq:form_z}) and Eq.~(\ref{eq:our_function}) are given in
column 5 of Table \ref{tab:z}. Note that this definition of $z_f$
allows for future (negative) formation redshifts.  The two definitions of
the formation redshift, $z_{1/2}$ and $z_{f}$, are correlated at $98
\%$ probability level (a Spearman rank correlation of $0.58$). In
addition, to evaluate the effect of the early and late portions of the
MAH on the formation redshift, we estimated the formation redshifts
using only the parts of the MAH for which 
 $a<0.65$ and $a>0.65$: $z_f^{<0.65}$ and
$z_f^{>0.65}$.  The best fit values of both $z_f^{<0.65}$ and
$z_f^{>0.65}$ are consistent with the values of $z_f$ within errors.
In addition the values of $z_f^{>0.65}$ are consistent with
$z_f^{<0.65}$ within (large) errors.  Although our cluster sample is
small, the significant spread in the values of $z_{f}$ for clusters of
the same $M_{180}$ is apparent. This indicates that halos of the same
mass exhibit a wide range of MAH shapes.
   
\section{Density Profiles}
\label{sec:density_prof}

\subsection{The fitting procedure} 
\label{sec:fit}

For each cluster halo, we fit the \citet[][NFW]{nfw_97},
the \citet[][M]{moore_etal98}, and the \citet[][JS]{jing_suto00}
analytic density profiles.  For a general profile of the form
\begin{equation}
\label{eq:nuker}
\rho(r)=\frac{\rho_{s}}{x^{\gamma} (1+x^{\alpha})^{(\beta - \gamma)/
\alpha}}, \ \ x\equiv r/r_{s},
\end{equation}
the NFW, M, and JS profiles have values of
($\alpha$, $\beta$, $\gamma$): (1, 3, 1), (1.5, 3, 1.5), and (1, 3,
1.5), respectively.

In what follows, we will define the concentration of a halo as
$c_{-2}\equiv r_{180}/r_{-2}$ and $c_{\rm v}\equiv r_{\rm
  vir}/r_{-2}$, where $r_{-2}$ is the radius where the logarithmic
slope of the best fit profile is equal to -2, $r_{180}$ is the radius
within which the average density is equal to 180 times the {\em mean}
matter density of the universe, and $r_{\rm vir}$ is the virial radius
defined using the redshift-dependent virial overdensity ($\approx 180$
at $z>1$ and $\approx 340$ at $z=0$).  To convert from
($M_{180}$,$r_{180}$) to ($M_{\rm vir}$, $r_{\rm vir}$) we use the
fitting formulas of \citet{hu_kravtsov03}.  For the general profile of
Eq.(~\ref{eq:nuker}), the radius $r_{-2}$ is given by
\begin{equation}
\label{eq:radius}
r_{-2}= \left( \frac{\gamma-2}{2-\beta} \right)^{1/\alpha} r_{s},
\end{equation}
where $r_s$ is the scale radius of the corresponding analytic profile.
Thus, $r_{-2}=r_{s}$ (NFW), $r_{-2} \approx 0.63 r_{s}$ (M), $r_{-2} =
0.5 r_{s}$ (JS).  In other words, $c_{-2}=c_{\rm NFW}$, $c_{-2} \approx
1.59 c_{\rm M}$, and $c_{-2}=2 c_{\rm JS}$, if the best fit is found to be the
NFW, the M, or the JS profile, respectively.  We present the resulting
concentration indices at $z=0$ in column 7 of Table~\ref{tab:z}.

There is a number of factors that may affect the fits of analytic
profiles to the profiles of the simulated clusters: the choice of
binning, the merit function, the range of radii used in the fitting, 
the weights assigned to the data points, etc. For example, for the merit
functions sensitive to the number of bins, such as $\chi^2$, the choice
of binning and bin weights are extremely important. 

In the following analysis, we use equal-size logarithmic bins in order
to give more statistical weight to the inner regions of the halos. The
number of bins is thirty for late epochs. For early epochs the number
of bins is reduced to ensure that each bin contains a sufficiently
large number ($>100$) of particles. We take as bin center the
average radius of all particles in a bin.  We checked that the
fits are robust when varying the number of bins around the adopted
value.  We find, however, that the choice of binning affects the
quality of the fit.  For example, for a large number of bins the
resulting profiles are quite noisy. Our choice of binning minimizes
the noise. We weight the data points by the poisson noise in the
number of particles of each bin.

The presence of substructure may substantially bias fits of smooth
analytic profiles. In particular, substantial amount of substructure
is present in the outer regions of halos and the profiles in these
regions are often non-monotonic exhibiting ``bumps''. To minimize the
bias, we fit the profiles using only the bins from a minimum resolved
radius (see $\S $~\ref{conver_study}) up to the radius within which
the average density is equal to 500 times the {\it critical} density
of the universe, $r_{500}$. This choice is motivated by the results of
\citet{evrard_etal96} who find that the material within this radius is
generally relaxed and in hydrostatic equilibrium.  We find that for
the clusters in our sample $r_{500}/r_{180}\simeq 0.36-0.37$ at $z=0$.
To the same end, the density profiles of CL9 and 10 were obtained by
averaging the $z=0$ and $z \approx 0.05$ outputs. This renders the
profiles less noisy and improves the quality of the obtained fits. The
averaging does not change the best fit parameters significantly. 

\begin{figure}[t]
\vspace{-1.cm}
\centerline{\epsfxsize3.8truein \epsfysize4.2truein\epsffile{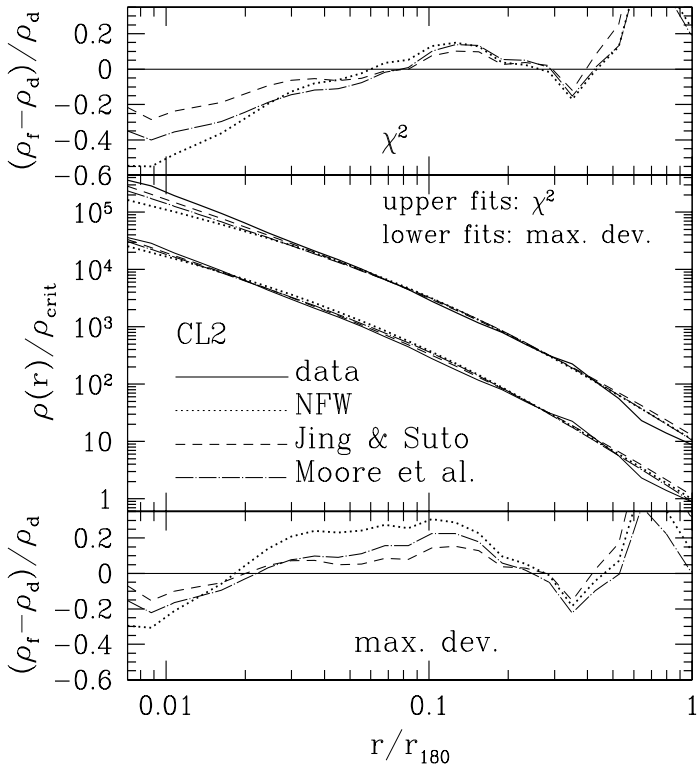}}   
\caption{{\it Middle panel:} NFW ({\it dotted lines}), Jing and Suto ({\it short-dashed lines}), and
  Moore et al. ({\it dot-dashed lines}) fits to the density
  distribution of CL2 ({\it solid lines}).  The upper set of lines
  corresponds to results obtained by a $\chi^2$ minimization and the
  lower set (displaced by a factor of 10 for clarity) to fits
  obtained by minimizing the maximum absolute fractional deviation of
  the fits from the data. {\it Upper panel:} Fractional deviation of
  the fits ($\rho_f$) from the data ($\rho_d$) for fits obtained via
  $\chi^2$ minimization. {\it Lower panel:} Same as in the upper panel
  but for fits obtained via maximum absolute fractional deviation
  minimization.
\label{fig:merit_function}}
\end{figure}

It is important to understand that the formal quality of the fit may
depend on the merit function, as well as the kind of binning and
weighting used.  In the following analysis, we fit the analytic
profiles for the parameters $\rho_{s}$ and $r_{s}$ by minimizing the
$\chi^2$.  \citet{klypin_etal01} show that $\chi^2$ merit function
applied to the profiles with logarithmic binning with the Poisson
error weights results in the fits of the NFW profile that are
systematically below the simulated profiles in the inner regions.  The
logarithmic binning gives higher density of data points at small
radii, creating thus bins with smaller number of particles.  The
$\chi^2$ fits for the CL2 profile are shown in Figure
\ref{fig:merit_function}.  Also shown are the fits obtained when using
the maximum fractional deviation (MFD) merit function, $\max[\vert\rho_{\rm
  fit}-\rho_{\rm data}\vert / \rho_{\rm data}]$, which gives equal
weight to all radial bins. Figure~\ref{fig:merit_function} shows that
this merit function reduces the deviations in the inner region
($r\lesssim 0.02r_{180}$) at the expense of significant deviations at
intermediate radii ($0.02\lesssim r/r_{180}\lesssim 0.2$).
Although the MFD merit function is less sensitive to the choice of 
binning, it is more sensitive to the presence of substructure 
bumps in the profile than $\chi^2$. Both merit functions have 
their pluses and minuses. 

Luckily, we find that regardless of the merit function and bin
weighting used, the relative goodness of fits for different analytic
profiles remains the same. If, for example, the M profile is a better
fit to the simulated profile than the NFW and JS profiles in the
$\chi^2$ minimization, it is the resulting best fit in the maximum
deviation minimization as well. The conclusion about which profile
fits best is therefore robust. Note, however, that this is not true
for the conclusions about the {\it systematic} ways by which a given
fit fails. For example, the characteristic 'S' shape of the fractional
deviation as a function of radius for the NFW fits found in various
studies \citep[e.g.,][]{moore_etal99b, ascasibar_03} is not a robust
result because it depends on the merit function, binning and
weighting.

In addition to the three widely used analytic profiles, we
experimented with fits of more general analytic expressions of the
form given by Eq.~(\ref{eq:nuker}). Overall, the fitting procedure
with all three parameters $\alpha, \beta, \gamma$ varied leads to
strong degeneracies between parameters \citep[see
also][]{klypin_etal01}. One can find several combinations of
parameters that fit the data equally well. For example, good fits with
inner asymptotic inner slopes as shallow as $\gamma=-0.3$ can be
found. 

In view of these degeneracies, we choose not to use generalized
fits but simply complement the analytic fits with measurements of the 
logarithmic slope profiles $s(r)\equiv
\partial \log\rho(r)/\partial\log r$. This analysis is complementary
to the fits because the slope is sensitive to the local shape of the
profile, while the fits may be sensitive to its global shape. The
logarithmic slope is computed using the linear fit to $\log\rho-\log
r$ locally. We use five neighboring profile bins centered on a given
bin in the fit (i.e., two bins on either side) with a total of 100
bins for the whole range of radii from $r_{min}$ to $r_{180}$.  The
choice of the number of bins is a tradeoff between the slope errors
and spatial resolution.  We experimented with fitting polynomials up
to the 4th order but found no advantage over a simple linear fit.  The
local logarithmic slope is sensitive to the presence of transient
massive substructures within the halo. For illustrative 
purposes only,  to reduce the
substructure-induced noise, we additionally smooth the slope using a
tophat filter. 

\subsection{Convergence Study}
\label{conver_study}

To study the effects of mass and force resolution, CL2 was
re-simulated with eight times more particles ($m_{\rm p}=3.95 \times
10^{7} h^{-1} M_{\odot}$) and with more refinements.  The ART code
performs mesh refinements when the number of particles in a mesh cell
exceeds a specified threshold. Thus, the mass resolution is tightly
linked to the peak spatial resolution achieved in simulation. The
cell size of the highest refinement level, which we will consider 
to be the formal resolution of the simulation, was $0.6h^{-1}$~kpc and
$1.2h^{-1}$~kpc in the higher- (HR) and lower-resolution (LR)
simulation, respectively. The HR simulation was initialized using the
same set of modes as the LR. We therefore follow the formation of the
same object with more particles. The comparison of density profiles
allows us to check for the two-body relaxation effects, which may be
important in cluster cores \citep{diemand_etal03}, and numerical
convergence.

\begin{figure}[t]
\vspace{-1.cm}
\centerline{\epsfxsize3.8truein \epsfysize4.2truein\epsffile{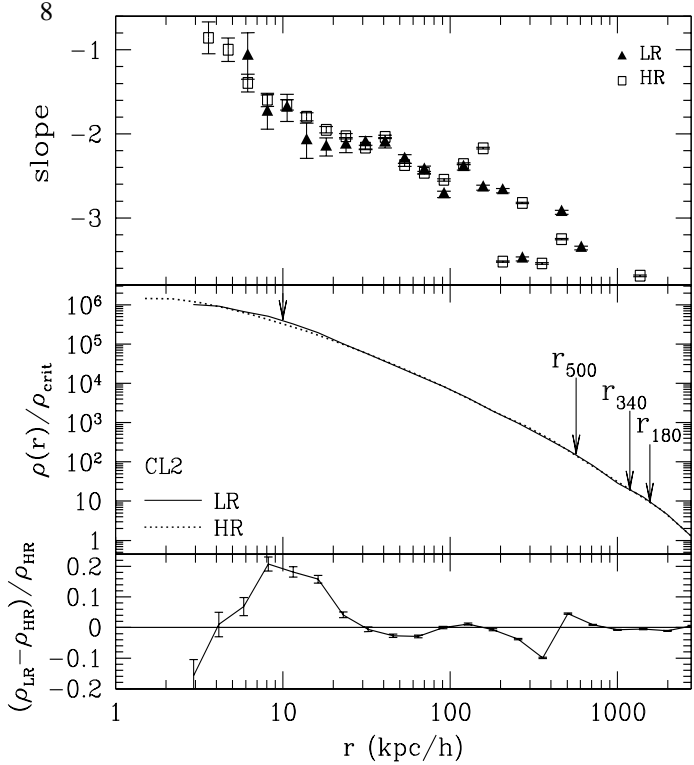}}   
\caption{
  Density profiles of the cluster CL2 in the low- (LR, {\em
    solid line}) and high-resolution (HR, {\em dotted line})
  simulations ({\em middle panel}). The profiles were obtained by
  averaging the $z=0, 0.02$ and 0.1 outputs of the corresponding runs.
  The {\it bottom panel} shows the fractional deviation between the LR
  and HR profiles. The error bars are computed by propagating the shot
  noise in the density profiles.  The {\it top panel} shows the local
  logarithmic slope as a function of radius in the HR ({\it squares})
  and the LR ({\it triangles}) runs.  In the middle panel, the
  vertical arrow at $\simeq 10\, h^{-1}$~kpc (or four times the formal
  resolution of the LR run) denotes the minimum distance used in our
  analyses. The arrows at large scales denote the radii corresponding
  to various commonly used overdensities: $r_{500}$, the radius within
  which the average density equals 500 the {\it critical} density, and
  $r_{340}$ and $r_{180}$, the radii within which the average density
  equals 340 and 180 times the {\it mean} density of the universe,
  respectively.
\label{fig:converg_1.000}}
\end{figure}

\begin{figure}[t]
\vspace{-1.cm}
\centerline{\epsfxsize=3.8truein\epsffile{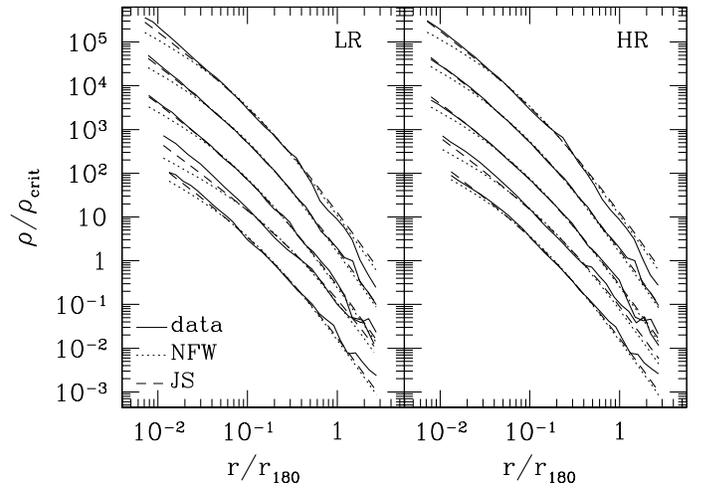}}   
\vspace{-1cm}
\caption{The density profile of CL2 at (from top to bottom) 
  $z=0$, $0.2$, $0.4$, $1$, and $1.5$ ({\it solid lines}), for the low
  (LR, {\it left panel}) and high (HR, {\it right panel}) resolution
  runs. The profiles at $z>0$ are scaled down by a factor of 10 with
  respect to each other. Also shown are the best fit NFW ({\it dotted
    lines}) and the JS ({\it dashed lines}) profiles. The figure shows
  that at all shown redshifts the JS profile is a better fit to the
  simulated profiles than the NFW in both runs.
  \label{fig:bestfit_evolution}}
\end{figure}

We compare the density profiles of CL2 in the HR and LR simulations in
Figure~\ref{fig:converg_1.000}. In order to minimize the differences
due to substructure, the profiles shown are obtained by averaging the
$z=0$, $0.02$, and $0.1$ outputs of the corresponding runs. The figure
shows that the fractional difference between the profiles is $\lesssim
0.2$ down to $\sim 3$ formal resolutions of the LR run.  This is in
agreement with a previous convergence study for the ART code using
simulations with lower mass resolution \citep{klypin_etal01}.
Comparison of density profiles of clusters in ART simulations with the
density profiles in simulations using the Gadget code
\citep{springel_etal01} was recently performed by
\citet{ascasibar_etal03}, who found excellent agreement between the
two codes at the resolved scales.

The upper panel of Figure~\ref{fig:converg_1.000} shows the local
logarithmic slope of the density profiles as a function of radius (see
\S~\ref{sec:fit} for details).
The error bars are computed by
propagating the Poisson errors in the density profiles.  At $r\gtrsim
200h^{-1}$~kpc the strong non-monotonic variations of the slope are due
to the presence of substructure. Despite the averaging, 
the small differences in the
locations of substructures result in large differences in the slope
value at a given $r$.  At the same time, the slopes in the HR and LR
runs agree well at scales $5\lesssim r\lesssim 200h^{-1}$~kpc. 
It is interesting to note that there is no evidence for a well-defined
asymptotic inner slope.  The local logarithmic slope in both runs
monotonically increases with decreasing radius down to the smallest
resolved scales.

Based on these results, in the subsequent analysis we will
conservatively consider only scales greater than $r_{\rm
  min}=10\,h^{-1}$~comoving kpc or (eight formal resolutions of the LR
run) for both the fits and the plots.  In addition, we require that
more than 200 particles are contained within the minimum radius
~\citep{klypin_etal01}.  In our simulations this criterion is relevant
only at early epochs ($z\gtrsim 2$), since at later epochs a
$10\,h^{-1}$ kpc radius always contains more than 200 particles for
all clusters.

One of the main results of our analysis is an apparent diversity of
the density profiles. At the same time, we find that if a profile is
best fit by a particular analytic profile at $z=0$, it is generally
best fit by the same analytic profile at early epochs out to $z\sim
1-2$.  Potentially, this is an interesting clue to the processes that
determine the shape of the profile. It is therefore important to check
that the conclusion does not change with resolution.
Figure~\ref{fig:bestfit_evolution} shows the fits of the NFW and JS
profiles to the profiles of CL2 in the LR and HR runs at different
epochs. The fits were done using bins in the radial range $[r_{\rm
  min},r_{500}]$ (see \S~\ref{sec:fit}).  The figure shows that at all
shown redshifts the JS profile fits the simulated profiles at small
radii better than the NFW profile in both runs. This is remarkable as
the cluster experiences fairly rapid increase in mass and several
violent mergers between $z=1.5$ and $z=0$.  The mass changes by more
than a factor of five during this period (see Fig.
~\ref{fig:growth_curves}). The cluster undergoes an intense merger
event at $z\sim 0.6$ and the last major merger for our definition
occurred at $z=z_{\rm LMM} \simeq 0.85$.

We find that our fitting results are robust to changes in both the
minimum and maximum radius used in the fits.  For example, concentration
changes by no more than $\simeq 10-20 \%$ if a different outer radius 
is used ($\sim 2/3 r_{180}$, and  for some clusters  an outer 
radius $\sim r_{180}$ did not change the results much). 
More importantly, the conclusion about the best fit analytic profile 
remains the same, although in some cases the best fit changes from the M
to the JS or vice versa. This can be expected because these analytic profiles
are quite similar.  We also repeated fits with twice as large minimum
radius ($\approx 20h^{-1} \ \rm{kpc}$ comoving) and find that 
the best fit analytic profile remains the same and that 
the concentration 
changes by $\lesssim 10\%$.

\subsection{Results}
\label{results}

\begin{figure*}[h!]
\epsscale{1.8}   
\plottwo{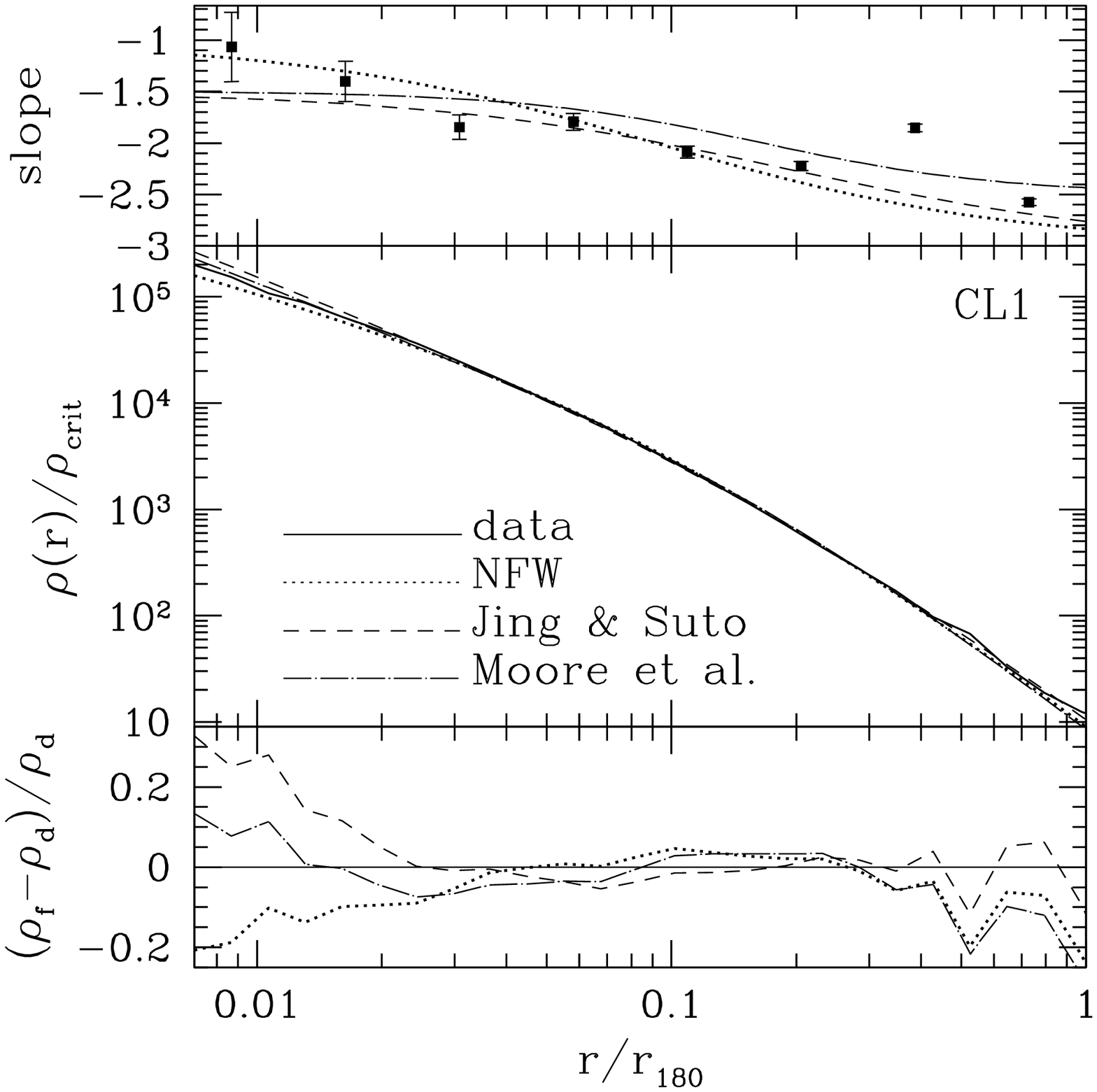}{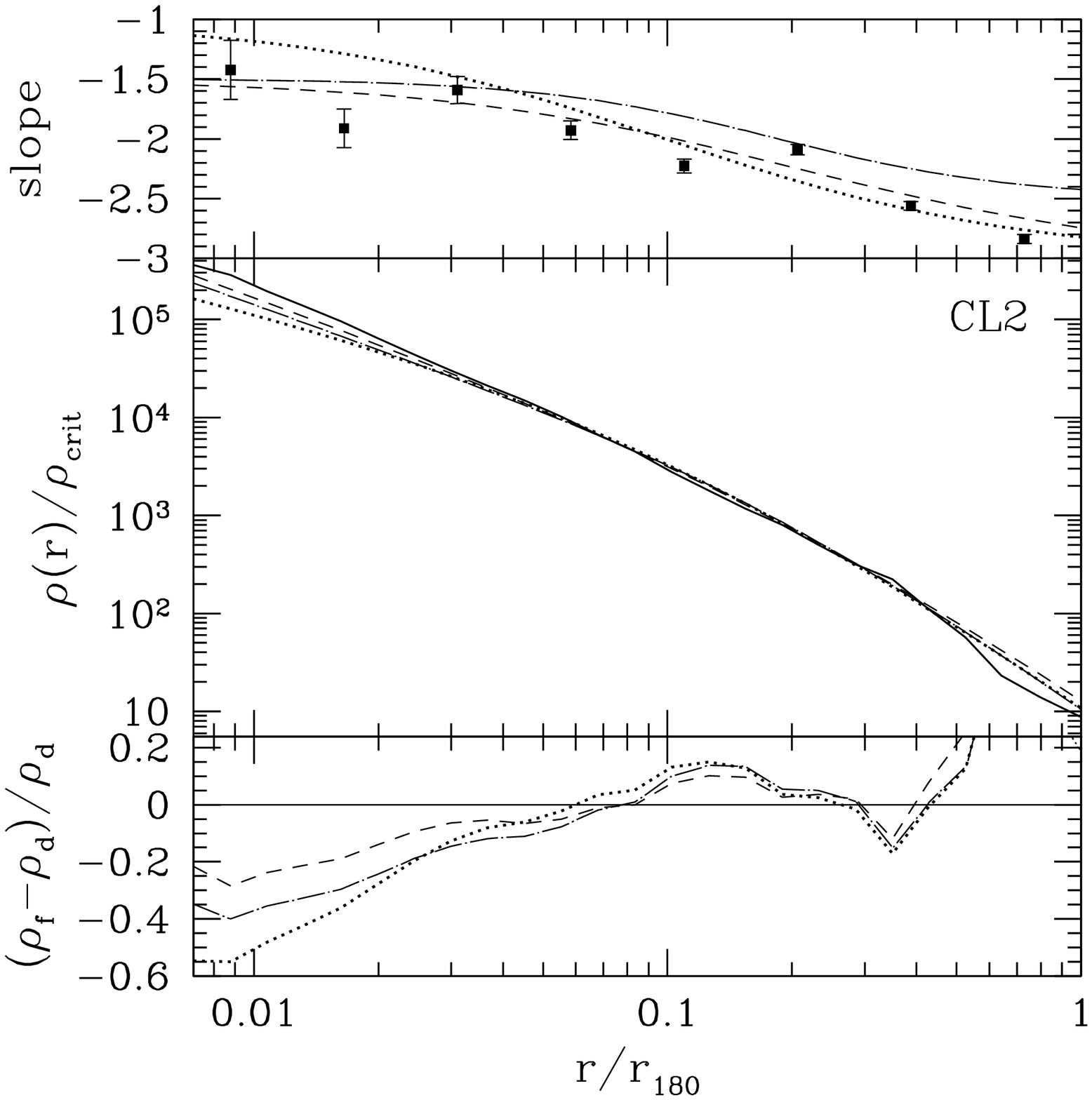}
\epsscale{1.8}
\plottwo{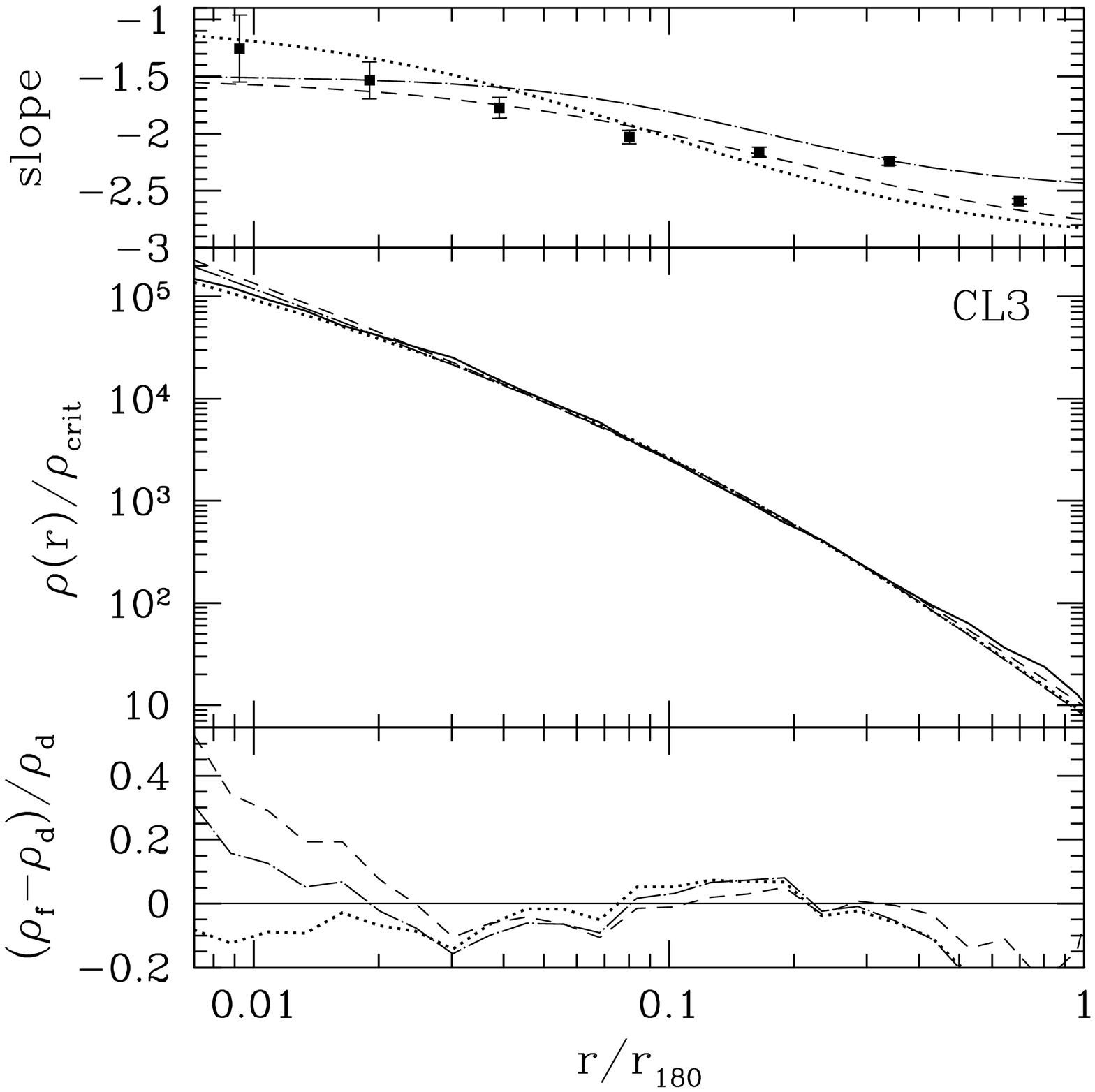}{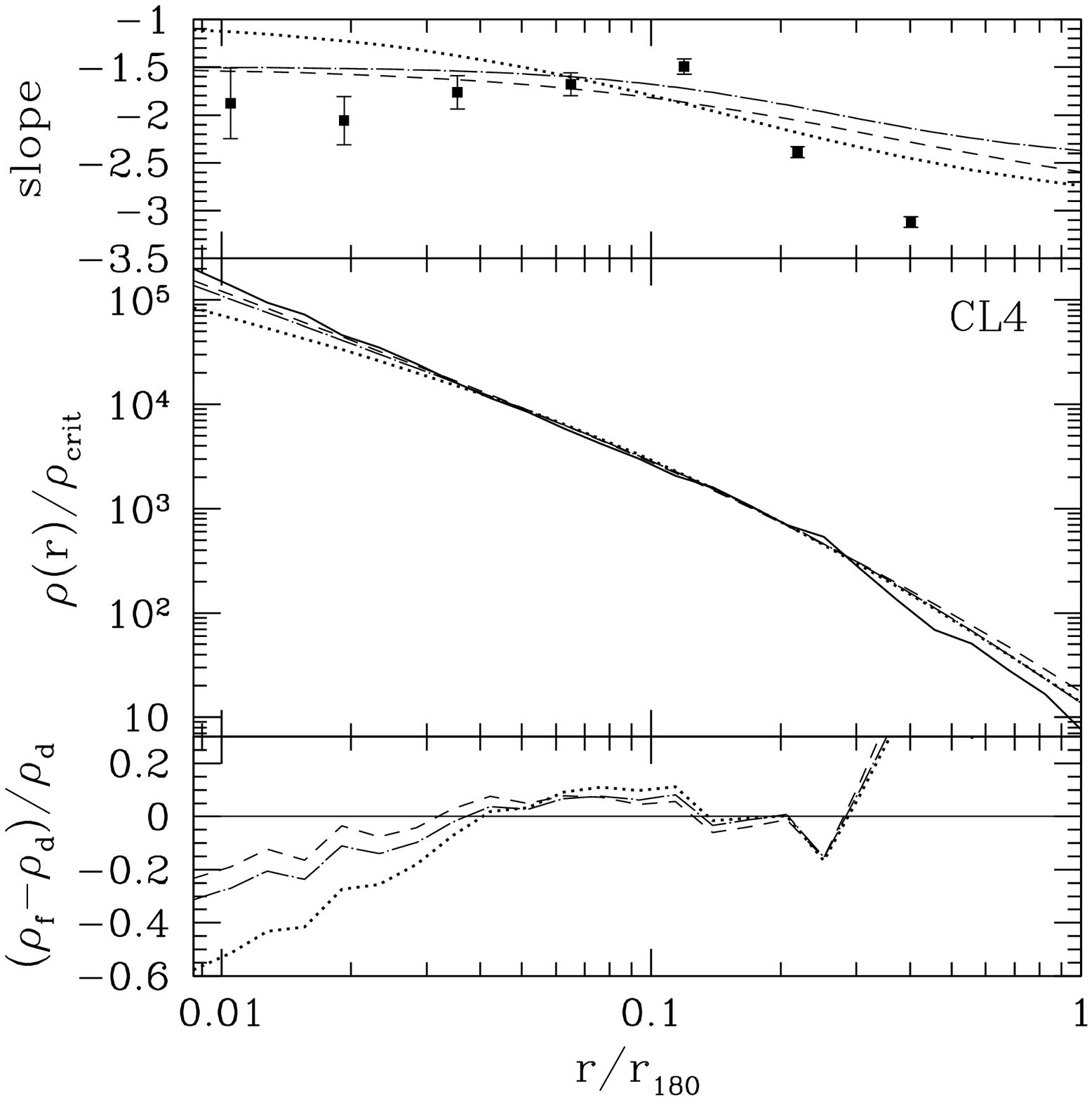}
\epsscale{1.8}
\plottwo{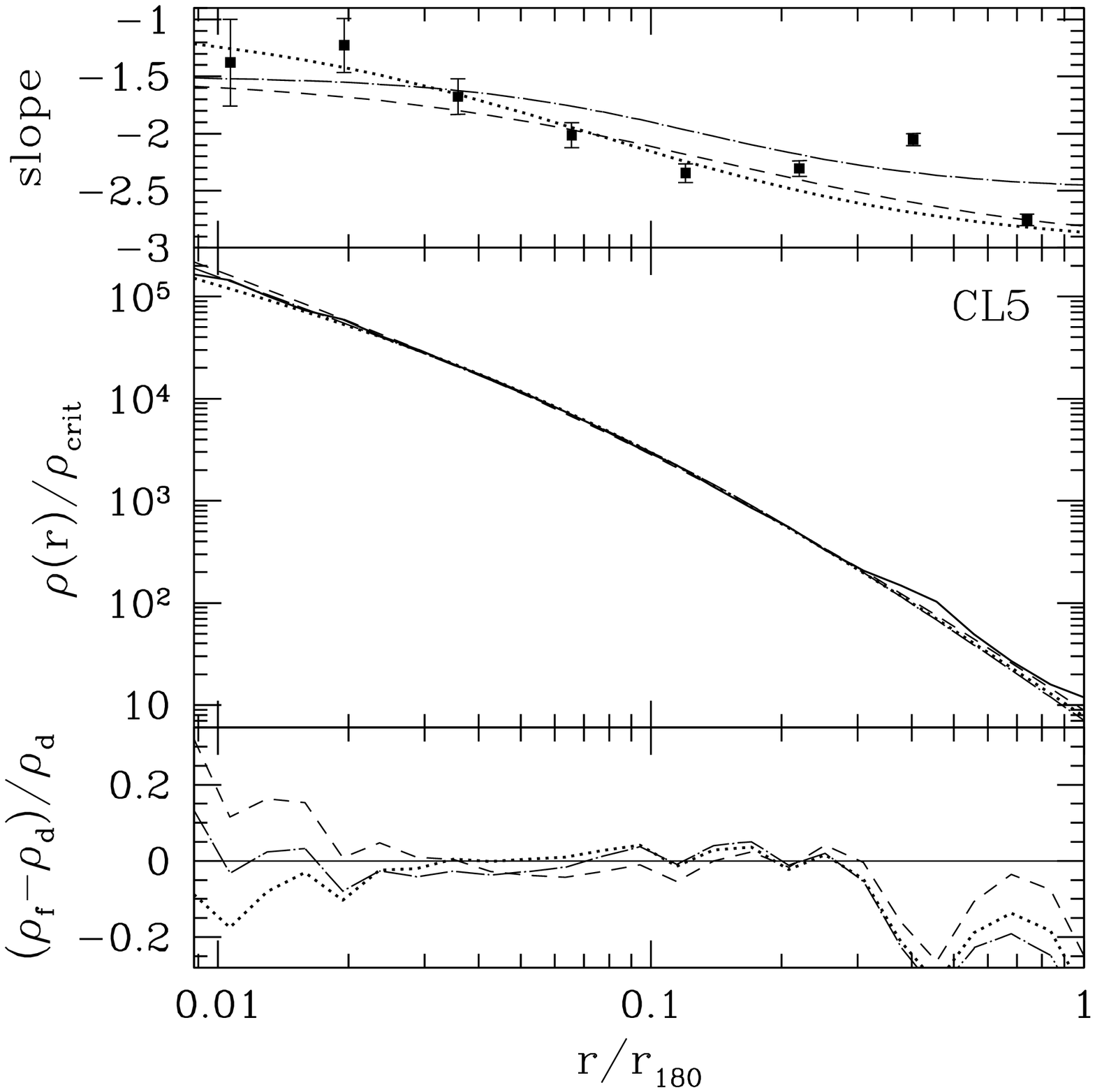}{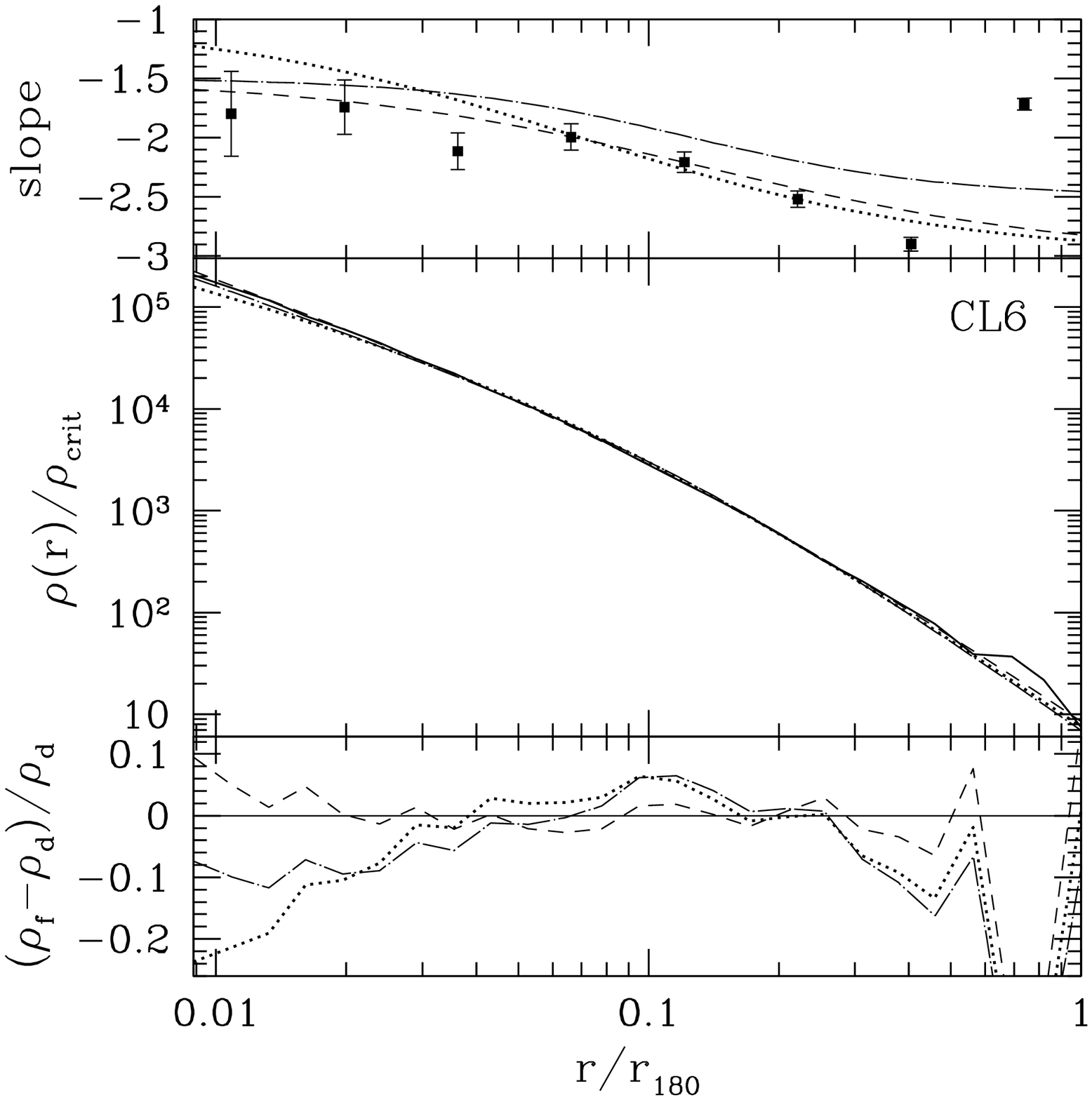}
\caption{{\it Middle panel:} fits of the NFW 
  ({\it dotted curves}), the Jing and Suto ({\it short-dashed
    curves}), and the Moore et al. ({\it dot-dashed curves}) profiles
  to the density distribution of the clusters of our sample at $z=0$.
  The fits were done using the range $[r_{\rm min},r_{500}]$ and a
  $\chi^2$ merit function (see \S \ref{sec:fit} for details).  The way
  the choice of merit function changes the fits can be seen in Figure
  \ref{fig:merit_function}.  {\it Bottom panels:} deviations of each
  one of the fits ($\rho_{f}$) from the actual profile ($\rho_{d}$).
  {\it Top panels:} local logarithmic slope as a function of radius
  for the 3 fits. The points correspond to the local slope as 
  derived from the actual profile. \label{fits}}
\end{figure*}

\begin{figure*}[h!]
\epsscale{1.8}   
\plottwo{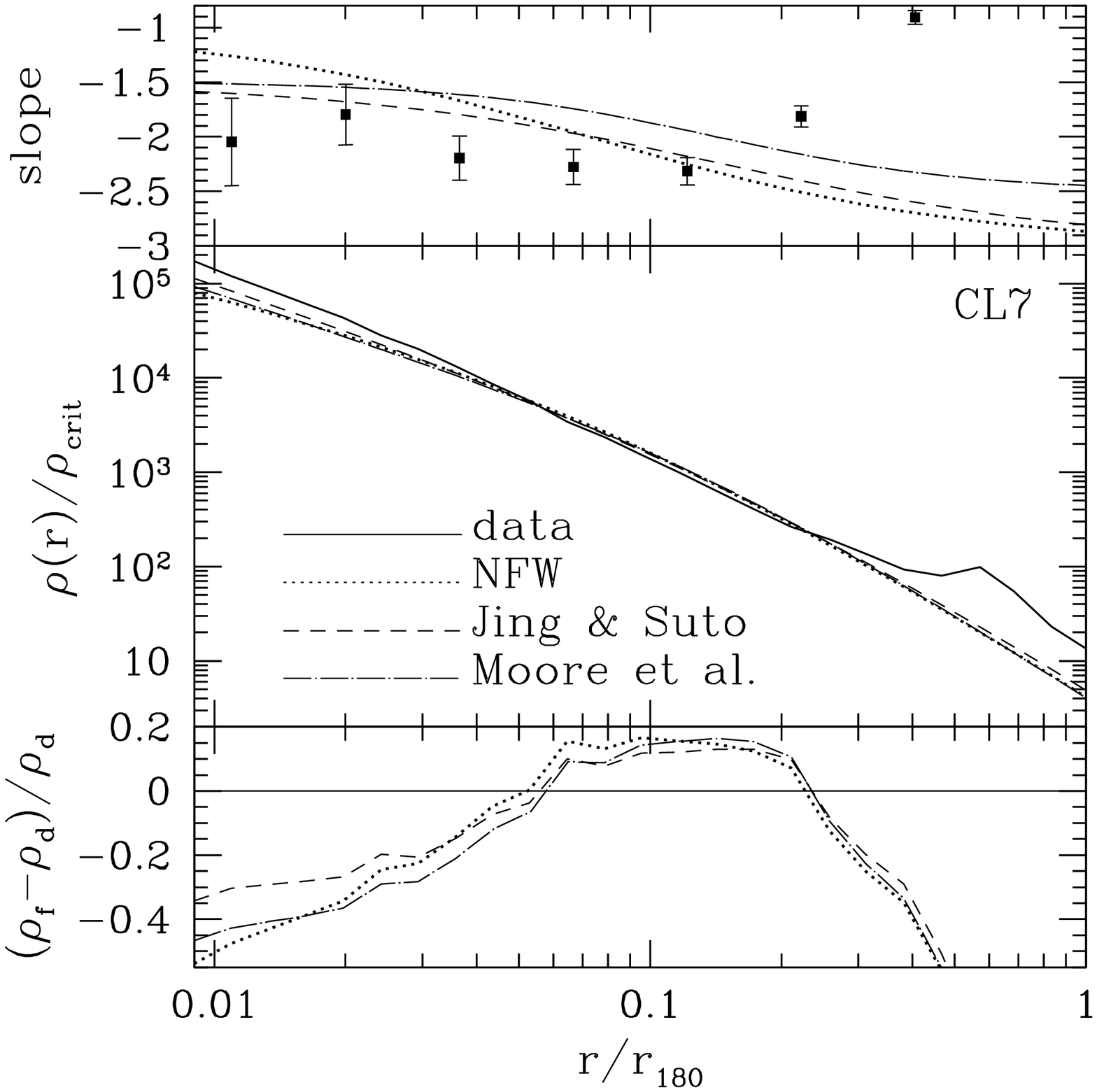}{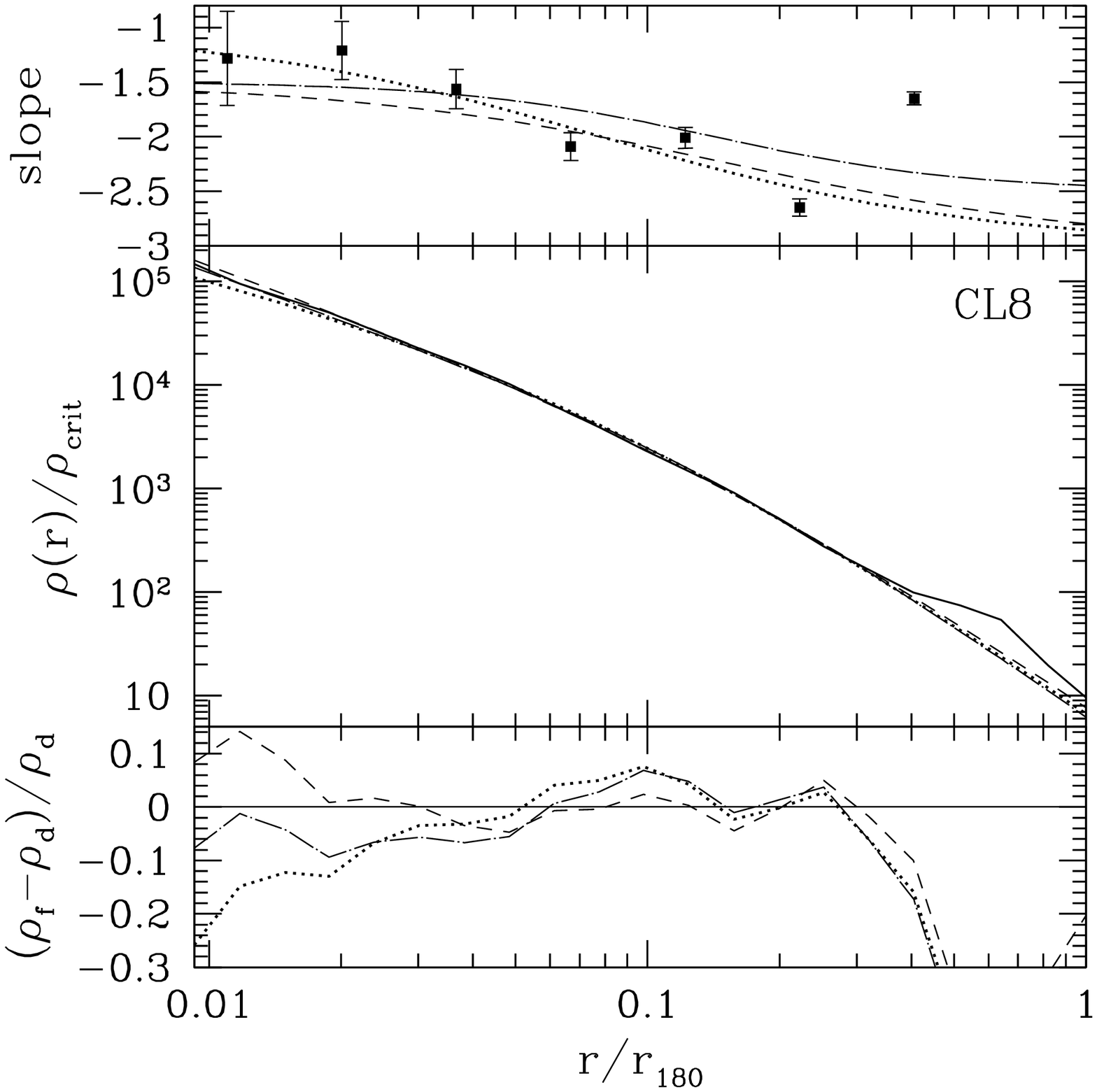}
\epsscale{1.8}
\plottwo{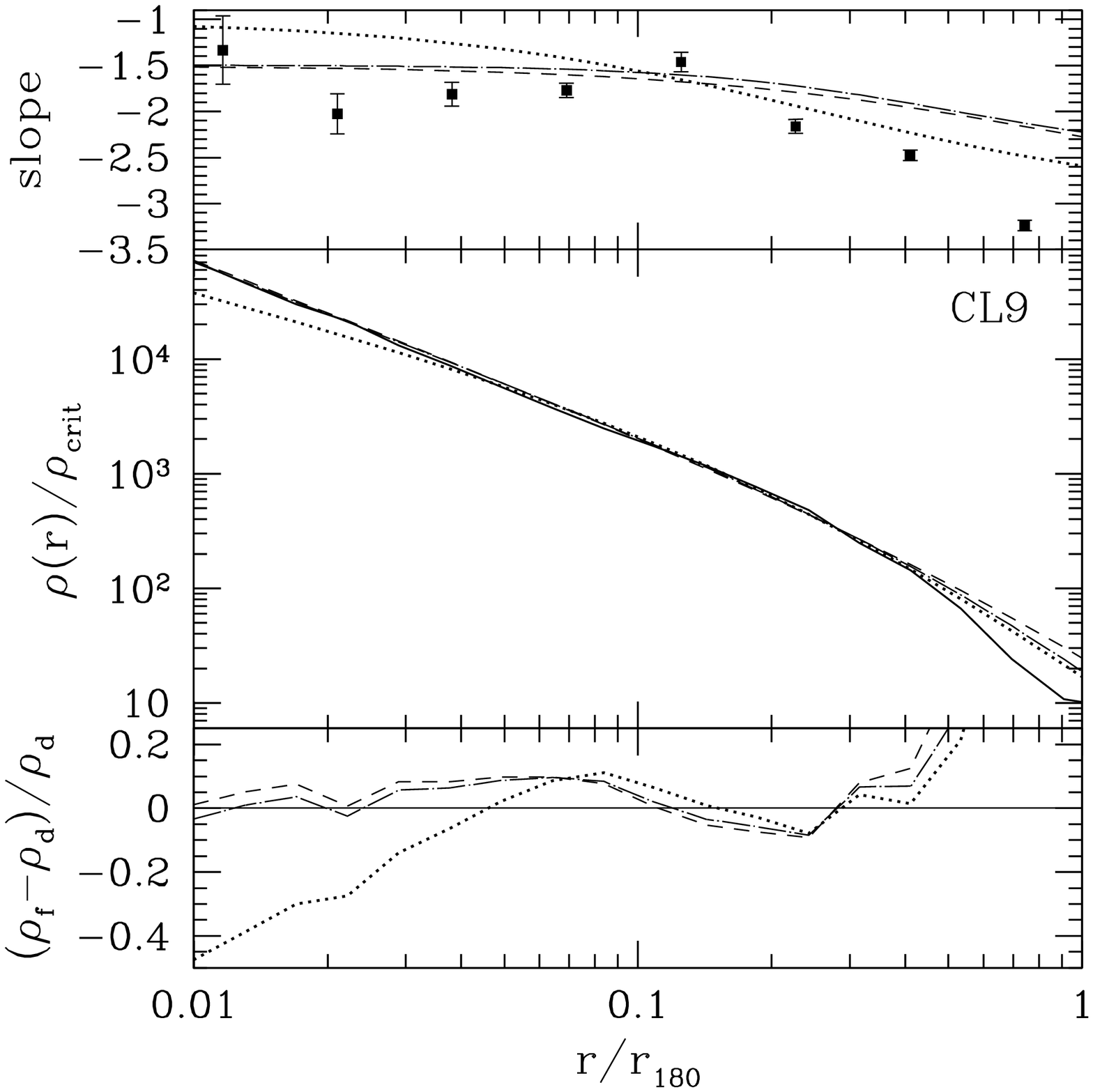}{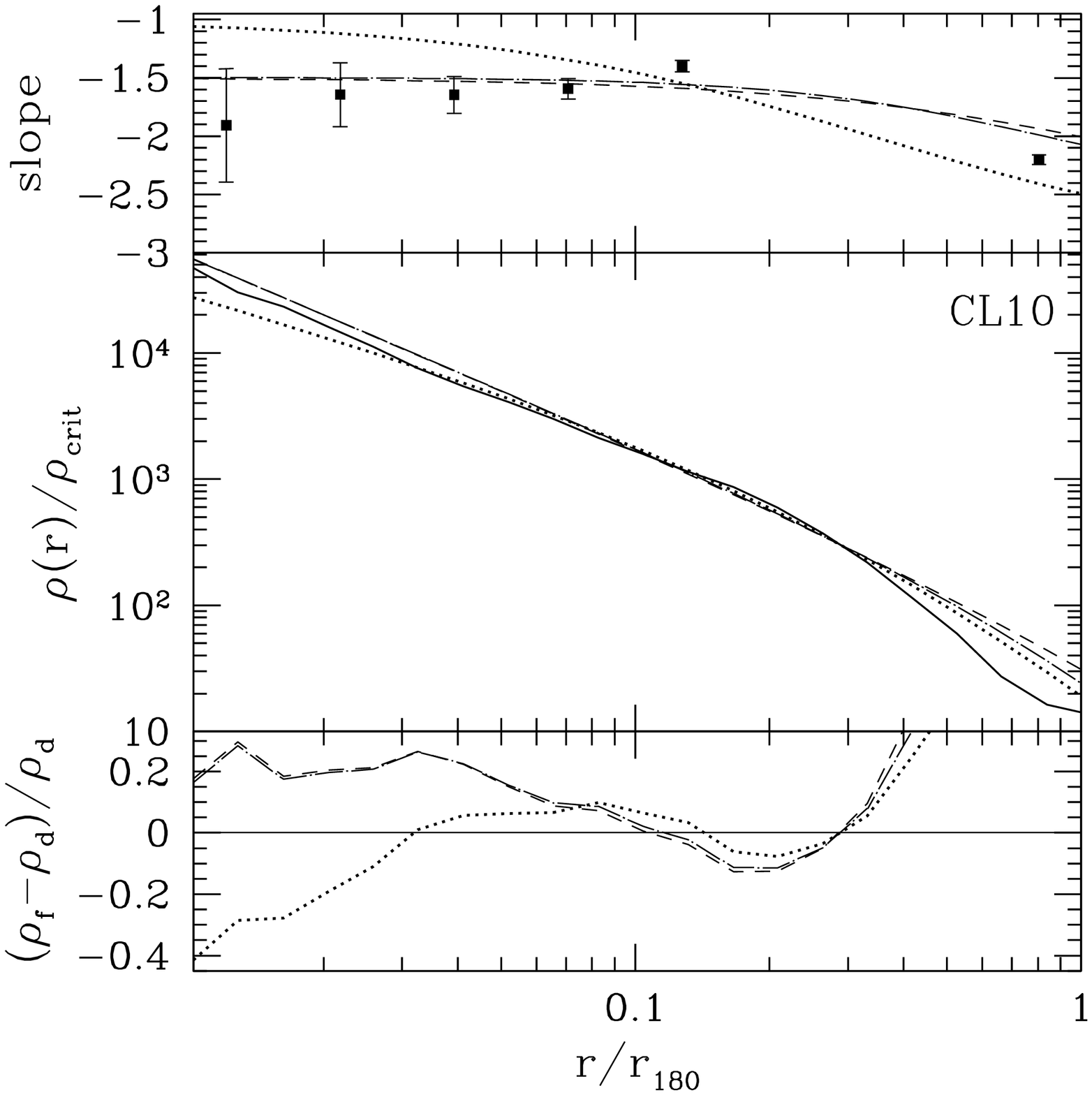}
\epsscale{1.8}
\plottwo{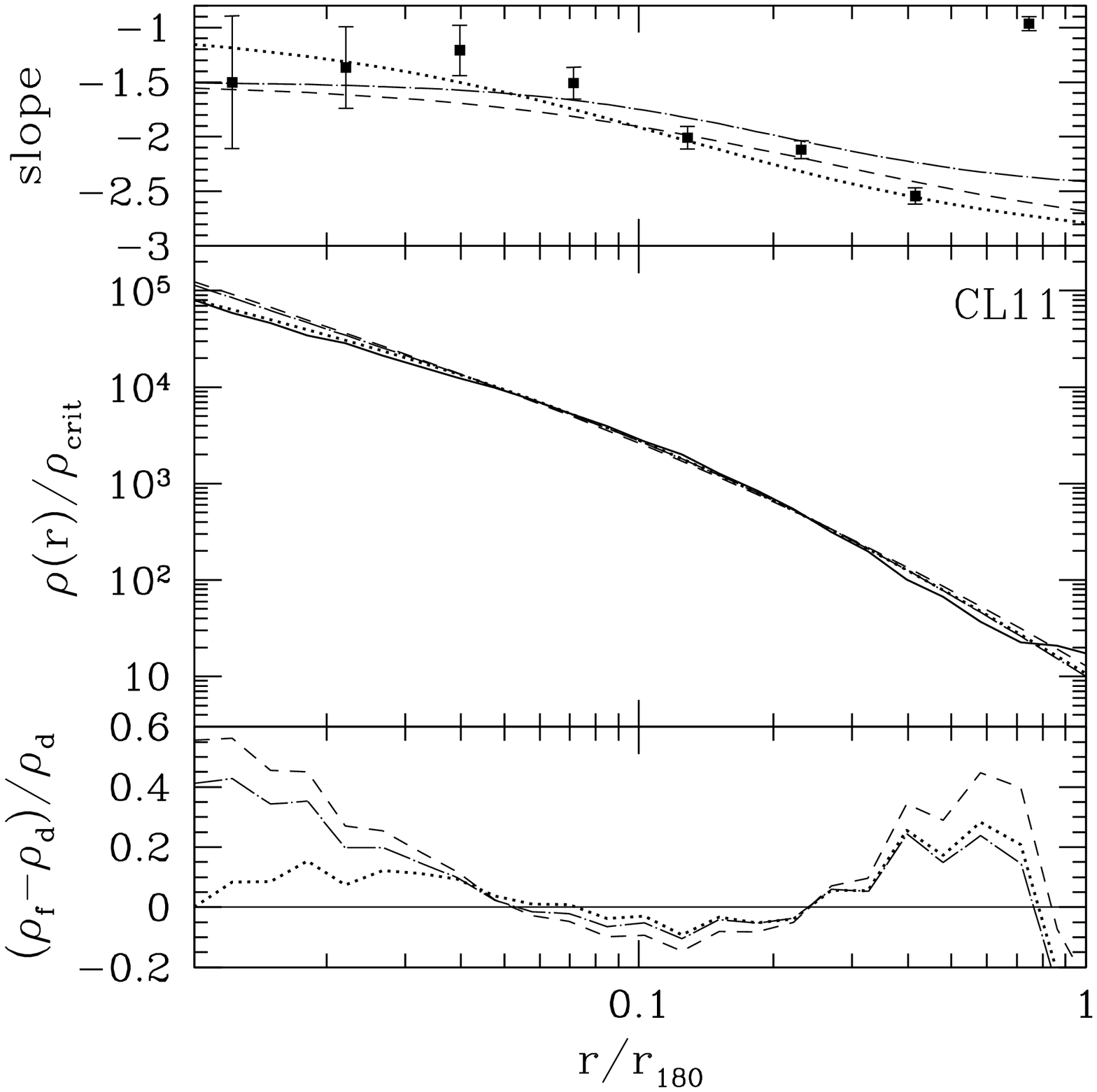}{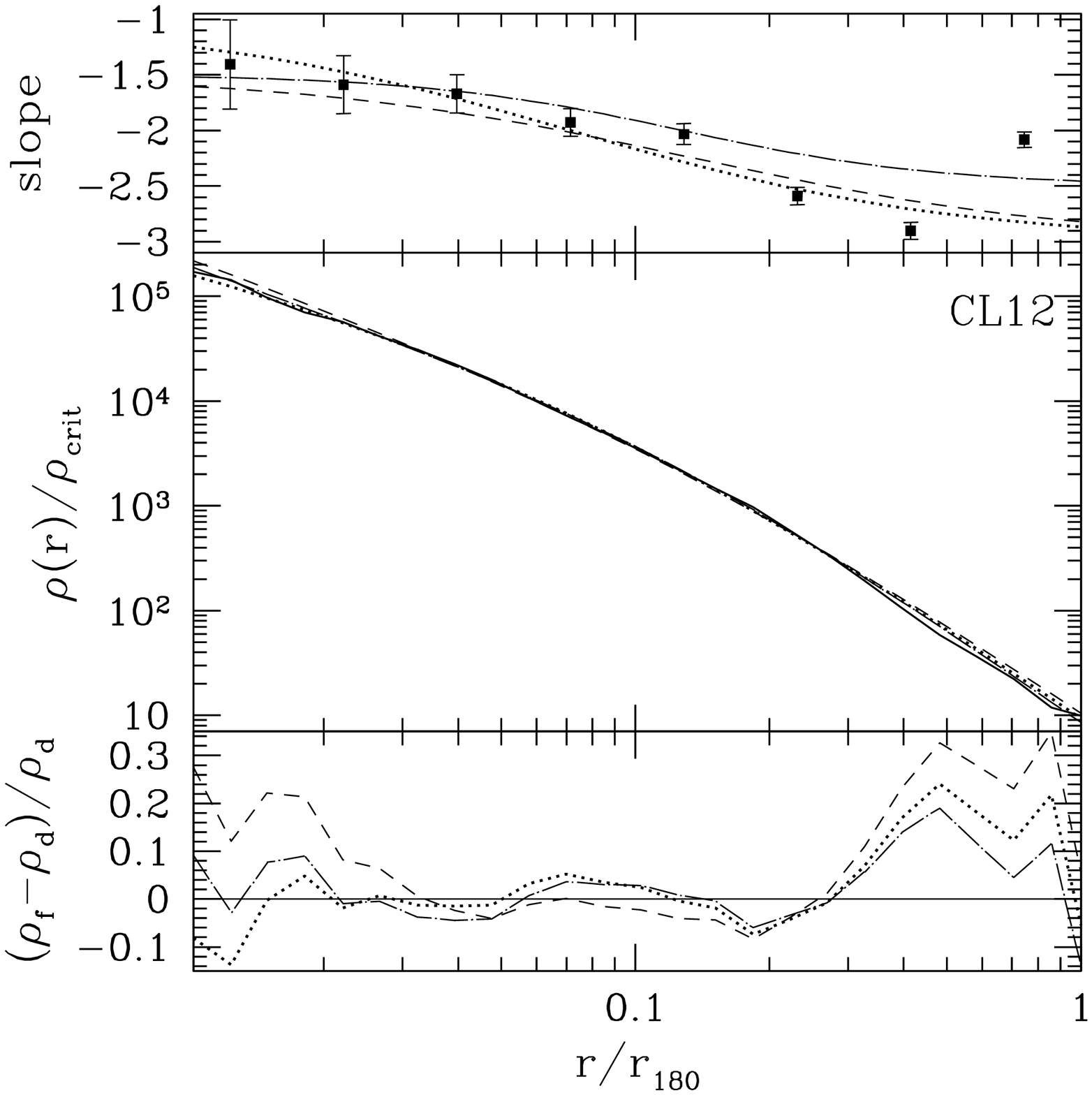}
\end{figure*}

\begin{figure*}[h!]
\epsscale{1.8}   
\plottwo{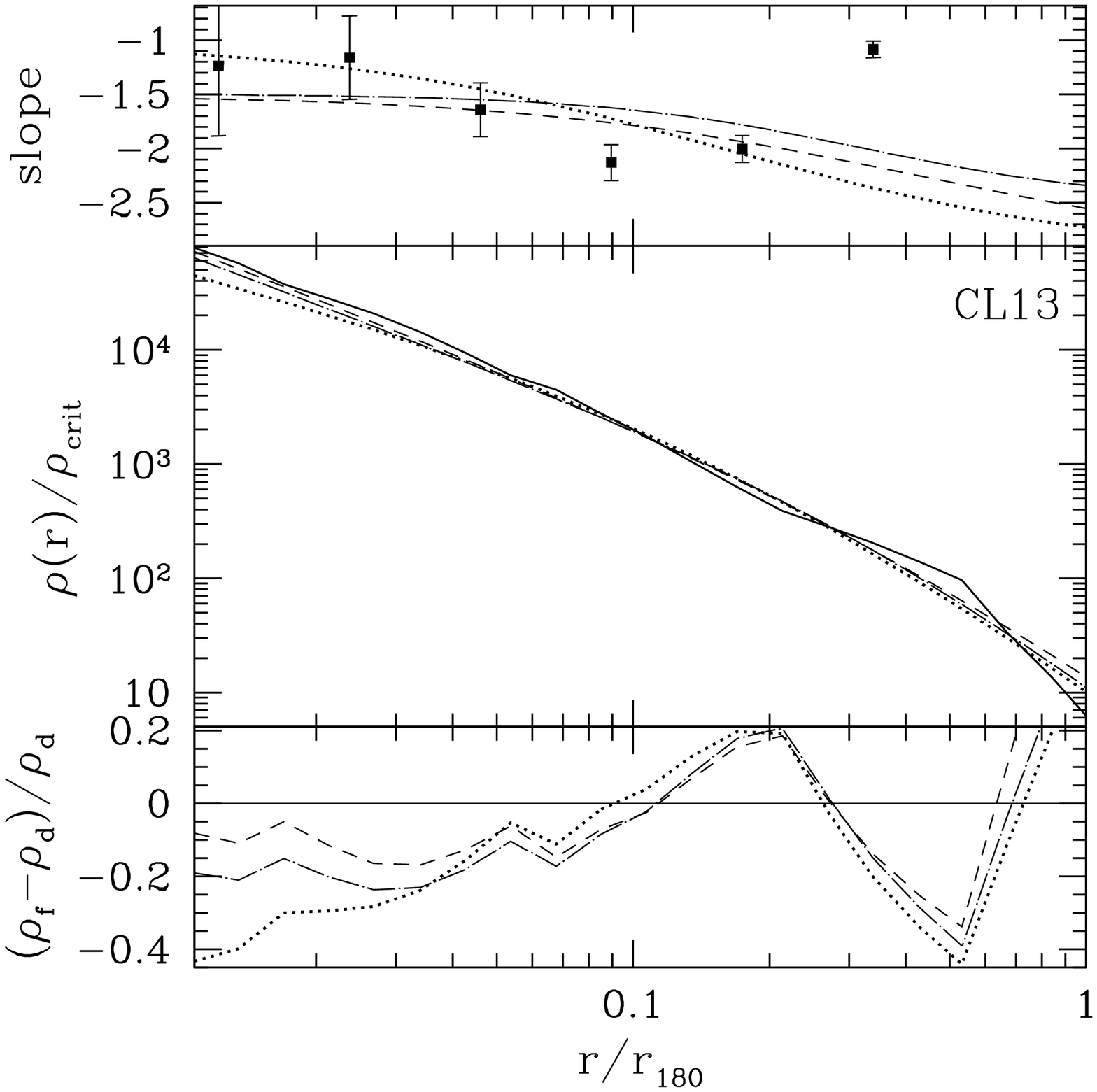}{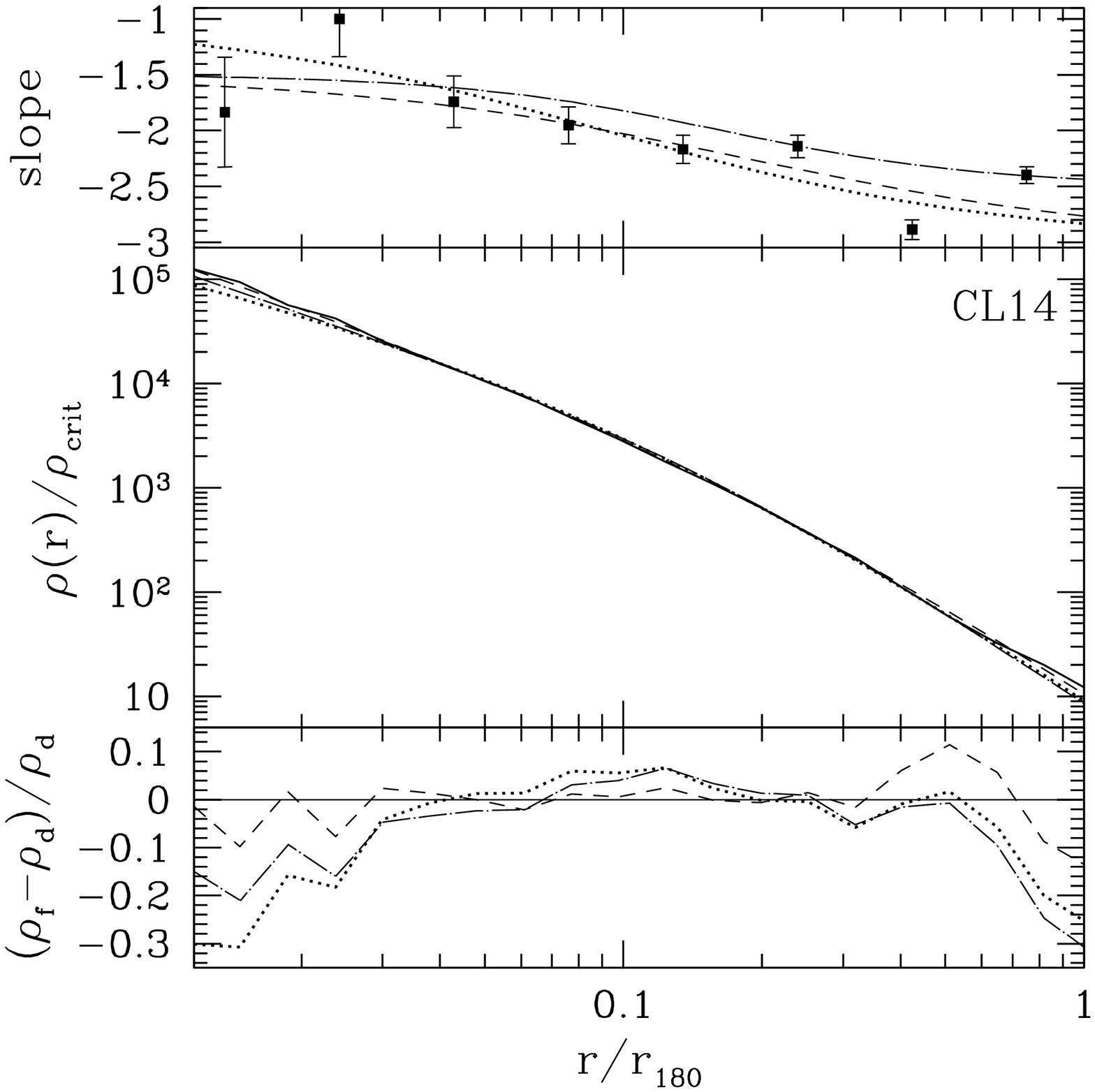}
\end{figure*}

Our results on the density profiles at $z=0$ are presented in
Figure~\ref{fits}.  In the middle panel we plot the actual profiles as
well as the NFW, M, and JS fits.  All profiles are plotted at
$r<r_{180}$ but are fit using only bins in the range $r_{\rm
  min}<r<r_{500}$, as discussed in $\S$~\ref{conver_study}.  In the
lower panel we present the fractional deviation as a function of
distance for each of the analytic fits. The analytic profile that provides
the best fit to the profile of each cluster is given in column 6 of
Table~\ref{tab:z}. In the upper panel we plot the logarithmic slope as
a function of radius for each of the analytic fits and the actual
local logarithmic slope as calculated from the simulated profiles (see
\S~\ref{sec:fit}).

In most cases the best analytic fit
provides by far the best fit compared to the other profiles.
This is especially true for CL3 and 11 (see Fig.~\ref{fits}), the two
clusters best fit by the NFW profile. The other two analytic profiles
fail significantly compared with the NFW for these clusters.
If we consider the similar M and JS as one family of profiles, the two
families (NFW and M/JS) typically differ significantly in quality.  As
discussed above, the systematic way in which the NFW profile fails to
fit the data can be attributed to the merit function used to obtain
the fits.  For our choice of merit function, the fits follow the
actual profile well at intermediate distances. The largest deviations
occur at the innermost regions.  For large distances, the three fits
are almost indistinguishable.

Figure~\ref{fits} and Table~\ref{tab:z} clearly show that there is
significant dispersion in the shapes of the profiles, concentrations,
and inner slopes. The dispersion of concentration parameter was
studied in several analyses
\citep{nfw_97,jing00,bullock_etal01,eke_etal01,wechsler_etal02,zhao_etal03a}
and is thought to be related to the distribution of the halo formation
epochs \citep{wechsler_etal02}.  The typical values of scatter are
$\sigma_{\log c_{v}}\approx 0.14$ with only a weak dependence 
on mass\citep{bullock_etal01,wechsler_etal02}.
Table~\ref{tab:z} shows that the $c_{-2}$ concentration indices 
of our clusters
span a wide range of values, from 2.3 to 14.7. Making the appropriate 
conversion from $c_{-2}$ to $c_{v}$, we find that the formal dispersion is
$\sigma_{\log c_{v}}\approx 0.2$ at $z=0$, larger than the dispersion for
the smaller mass halos used to derive this scattering in other
studies, which may reflect the more recent formation times of cluster
halos, as well as their more diverse MAHs.  For example,
\citet{klypin_etal03} and \citet{colin_etal03} find a significantly
smaller dispersion $\sigma_{\log c_{v}}\approx 0.1$ for a subsample of
relaxed halos without significant substructure. Indeed, the clusters
with the three lowest concentrations, CL9, 10, and 13 have all had a
recent major merger (see $z_{LMM}$ in column 2 of Table~\ref{tab:z}).
The small concentration of these objects is due to the shape of their
density profiles which are close to a power law over a wide range of
radii.  Careful examination of MAHs and merger histories, indicates
that the recent major merger activity results in a low concentration
of density profiles. CL4, which also has a small concentration
compared to the majority of the clusters, has a formal $z_{LMM}=0.95$,
for the definition of major merger adopted in our study. However, in
agreement with the other low concentration objects, it had a large
merger ($\simeq 22 \%$ fractional mass increase) at a very recent
epoch ($z \simeq 0.15$).  In addition, we find a strong correlation
between $z_{f}$ and $c_{-2}$, in agreement with the correlation
advocated by W02.

\begin{figure}[t]
\centerline{\epsfxsize=3.5truein\epsffile{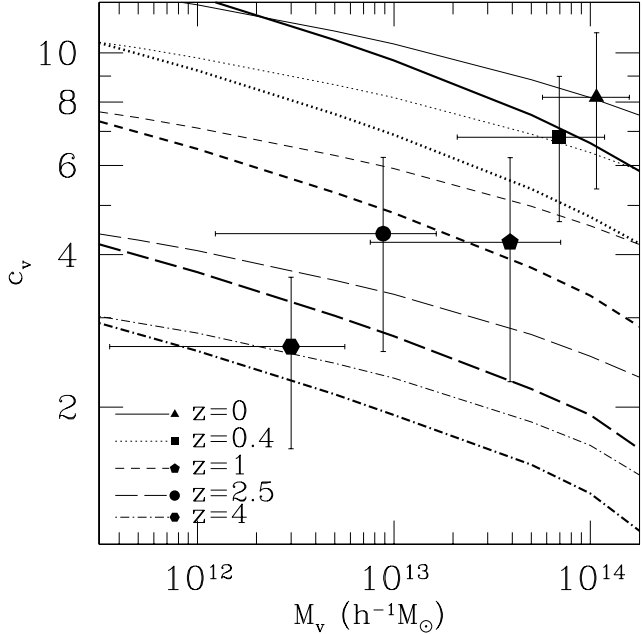}}   
\caption{Median concentration vs. virial mass at different
  redshifts for the progenitors of clusters in our sample ({\it
    points}). The vertical errorbars represent the 1$\sigma$ scatter in
  concentrations for the 14 clusters, while horizontal errorbars show
  the mass range of the halos at each epoch. The predictions 
  of the \cite{bullock_etal01} ({\it thick lines}) 
  and the \cite{eke_etal01} ({\it thin lines}) models are plotted for comparison. 
\label{fig:cv_mv}}
\end{figure}

Figure \ref{fig:cv_mv} shows the redshift evolution of the median
virial concentration, $c_{v}$, of our sample. We also plot the
predictions of the models by \citet{bullock_etal01} and 
\citet{eke_etal01}. Our results seem
to be in some general agreement with both model predictions. 
Overall, the \citet{eke_etal01} seems to be in better
agreement  than the \cite{bullock_etal01} model. 
Recently, \citet{dolag_etal03} found that
the standard \citet{bullock_etal01} recipe systematically
underestimates concentrations of cluster-size halos at all redshifts.
Figure \ref{fig:cv_mv} shows a similar trend in our simulations,
although the difference we find is noticeably smaller.  This may be
due to the smaller mean mass of clusters in our sample.
\citet{zhao_etal03a}, for example, show that discrepancy between
simulations and the \citet{bullock_etal01} model increases with increasing
halo mass.  In the mass range probed here the difference from the
analytic prescription of \citet{bullock_etal01} is considerably
smaller than the scatter in concentrations.

\begin{figure}[t]
\centerline{\epsfxsize=3.5truein\epsffile{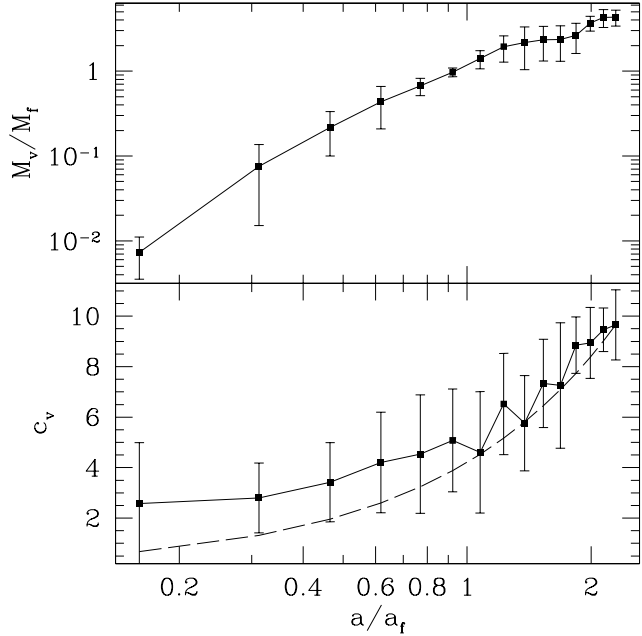}}   
\caption{Average mass accretion history  ({\it top panel}) 
  and average concentration of cluster progenitors ({\it bottom panel}) as a
  function of scale factor measured in units of the formation scale factor, $a_{\rm f}$. 
  The errorbars in both panels represent the
  1$\sigma$ spread around the mean.  
   The figure shows that
 the concentration is approximately constant at $c_{\rm v}\approx 3-4$ 
during the period of rapid mass accretion ($a/a_{\rm f}<1$) 
and increases with decreasing redshift during the period of slow mass 
growth ($a/a_{\rm f}>1$). For comparison the {\em dashed line} shows 
$(1+z)^{-1}$ evolution. 
\label{fig:cmv}}
\end{figure}

Figure \ref{fig:cmv} shows the evolution of the average concentration
and the average MAH of our clusters. The figure shows that during the
period of rapid mass growth the concentration is nearly constant at
$c_{\rm v}\approx 3-4$, while during the period of gradual mass growth
it increases with decreasing redshift as $c_{\rm v}\propto
(1+z)^{-1}$. Therefore, the concentration of halos is approximately
constant while they experience frequent major mergers and there may
exist a ``floor'' to the concentration values of $c_{\rm min}\approx
3$, while concentrations start to increase with time at $z<z_{\rm f}$.
This behavior was pointed out by \citet{zhao_etal03a,zhao_etal03b}.
However, unlike \citet{zhao_etal03a}, we find that the model of
\citet{wechsler_etal02} does not underestimate the concentrations at
the cluster-size halos of our sample. The evolution of average
concentration in Figure~\ref{fig:cmv} is similar to that found by
\citet{dolag_etal03}.  The mean concentration of cluster progenitors
in their simulations (their Figures~4 and 7) is approximately constant
at $z\gtrsim 1$.

\begin{figure}[t]
\vspace{-1.cm}
\centerline{\epsfxsize=3.8truein\epsffile{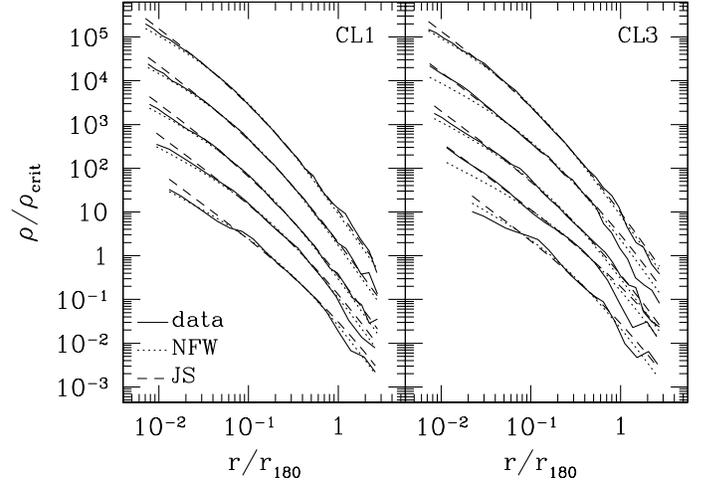}}   
\vspace{-1cm}
\caption{The density profile of CL1 ({\it left panel}) and 
  CL3 ({\it right panel}) at (from top to bottom) $z=0$, $0.2$, $0.4$,
  $1$, and $1.5$ ({\it solid lines}).  The profiles at $z>0$ are
  scaled down by a factor of 10 with respect to each other. Also shown
  are the best fit NFW ({\it dotted lines}) and the JS ({\it dashed
    lines}) profiles. At all shown redshifts the NFW profile is a
  better fit to the simulated profile of CL1 than the JS profile. CL3
  is shown as an exception to the typical behavior represented by CL1
  and CL2 (Fig.~\ref{fig:bestfit_evolution}).  For CL3 the best fit at
  $z=0$ (NFW) is not the best fit at earlier epochs.
\label{fig:bestfit_CL7and5}}
\end
{figure}

We do not find a clear connection between the redshift of last major
merger and the best fit profile (column 6 in Table~\ref{tab:z}).  We
do find that none of the objects with low $z_{LMM}$ has the NFW
profile as the best fit, but the statistics of our sample is too small
to reach a firm conclusion. Nevertheless, for each individual system
the shape of its density profile is remarkably stable during
evolution.  The best fit analytic profile at $z=0$ is typically also
the best fit at earlier epochs, as was shown for CL2 in
Figure~\ref{fig:bestfit_evolution}.  Figure~\ref{fig:bestfit_CL7and5}
shows evolution of the density profiles and the analytic fits at each
epoch for CL1 and CL3. The NFW profile is a better fit than the JS at
all epochs for $z<1.5$ for  CL1.  This stability of the
profile shape with time holds for most clusters with some
exceptions. CL3 illustrates the case where the best fit analytic
profile changes from epoch to epoch. We find this behavior
for four out of the fourteen clusters in our sample.

Note that the NFW fits never reach their inner asymptotic slope of -1
at the radii we probe. The average logarithmic slopes estimated by 
averaging  the local slope 
around  $0.03r_{180}$ (see \S \ref{sec:fit})
range from $-1.2$ to $-2$, and are given in column 8 of Table~\ref{tab:z}. 
In most cases, the local slope changes
monotonically with radius with no sign of reaching the asymptotic
inner slope. Note that for typical concentrations of cluster halos we
expect the asymptotic slope to be reached at the resolved scales, at
least for the M profile\footnote{From Eq.~(\ref{eq:nuker}) $r_{s}$
  is the radius where the logarithmic slope is equal to $-(\beta +
  \gamma)/2$ with the asymptotic slope reached at $r << r_{s}$. The
  NFW and the JS profiles have typically smaller $r_{s}$ than that of
  the M profile. As a result, they reach their asymptotic slopes at
  smaller distances. An analytic profile, of course, can be a good fit
  regardless of whether its inner asymptotic slope is resolved.}. This
can be seen from the slope profiles for the best fit analytic fits
shown in the top panels of Figure~\ref{fits}. The only cases where the
slope profiles are flat for  relatively large radius ranges, 
correspond to the halos with
recent merger activity and rapid mass growth (i.e., CL4, CL9, CL10,
CL13).  

Interestingly, we find that the density profiles of systems that
experience intense merger activity until the present epoch (clusters
CL4, 9, 10 and 13) can be well described by a single 
power law $r^{-\gamma}$
with slope $\gamma$ ranging from -1.5 to -2. Similarly to other
profile shapes, the power law density profile for these systems is
maintained for earlier epochs out to $z\sim 1.5$.  In addition, there
is evidence that the profiles of all clusters during their rapid mass
growth stages are close to a power law. In particular, we find that
the power law provides an increasingly better fit with increasing
redshifts for all of our clusters. At early epochs ($z\gtrsim 1.5-2$),
the power law fit is always either comparable or better than the NFW,
M, and JS analytic profiles. The power law like profiles 
relate to  low concentrations. This is because they maintain a
slope slightly shallower than $-2$ out to large radii so that the
scale radius is large. Indeed, the clusters with power law profiles at
$z=0$ have the lowest concentrations in our sample. The decrease of
concentrations at higher redshifts may thus reflect the power law like
density profiles of actively merging systems.

\section{Discussion and conclusions}
\label{sec:conclusions}

We have studied the MAHs and density profiles of 14 cluster-size halos
simulated using the ART code in a flat $\Lambda$CDM cosmology. In
agreement with previous studies, we find that most MAHs have a similar
shape: an early, merger-dominated mass increase followed by a more
gradual, accretion-dominated growth.  To obtain a formation redshift
that characterizes the overall shape of the MAH we perform analytic
functional fits. The typical MAHs are well described by the one
parameter exponential function proposed by W02
(Eq.~[\ref{eq:w02mah}]).  Two of the clusters in our sample experience
intense merger activity and rapid mass growth until the
present-day epoch. The MAHs of these systems are better described by a
one parameter power-law function in the scale factor. We
thus generalize the form proposed by W02 into a two parameter form
(Eq.~[\ref{eq:our_function}]) to encompass both exponential and
power-law MAHs.  For each class, however, the fit reduces to a one
parameter fit.

We check the convergence of halo density profiles using a
re-simulation of one of the clusters with eight times more particles
and better force resolution.  We show that both the halo profiles and
their local logarithmic slopes converge at scales larger than
about four times the formal resolution of the low resolution run, in
agreement with a previous convergence study of the ART code by
\citet{klypin_etal01}. We fit the density distribution of the clusters
with the NFW, M, and JS analytic profiles. Experiments show that the
choice of merit function, weighting, and binning affect the absolute
quality of a fit and may bias conclusions about how well a particular
analytic profile fits simulation results.  We find, however, that the
relative goodness of fit for the three analytic profiles and our
conclusions about the best fit profile are robust to the changes in
binning and merit function.

The main general result of our study is a remarkable diversity of the
mass accretion histories, profile shapes, concentrations, and inner
slopes for cluster-size halos in a relatively narrow mass range.  The
concentrations of cluster-size halos at the present-day epoch exhibit
a scatter of $\sigma_{\log{c_v}}\approx 0.20$. This scatter is related
to the diversity of halo MAHs and formation redshifts.  We find a
statistically significant correlation between the formation redshift
and the concentration of a halo, in agreement with results of W02.
There is a more detailed connection between the MAH and concentration.
The concentration of a halo is approximately constant at $c_{\rm
  v}\approx 3-4$ during the period of rapid mass growth and frequent
major mergers ($z>z_{\rm f}$) and increases with decreasing redshift
when the mass accretion rates slows down at $z<z_{\rm f}$. This
behavior was recently pointed out by
\citet{zhao_etal03a,zhao_etal03b}. The implied ``floor'' in the
concentration is not accounted for in the currently used models for
$c_{\rm v}(M)$ \citep{bullock_etal01,eke_etal01}, which predict a
monotonic decrease of concentration with increasing mass and may thus
underestimate concentrations of the most massive, $\gtrsim 5\times
10^{14}h^{-1}\ \rm M_{\odot}$, halos.  This may have important
implications for estimates of expected number of wide-separation
quasar lenses \cite[e.g.,][]{kuhlen_etal03} and other results
sensitive to the concentrations of very massive clusters.

The inner logarithmic slope of cluster profiles at $3\%$ of the virial
radius (or $10-50$~kpc) ranges from $-1.2$ to $-2$. In the best
resolved clusters the logarithmic slope does not seem to reach a
specific asymptotic value down to the smallest resolved scales in our
simulations ($r/r_{180} \simeq 0.007$). A similar conclusion was
reached by \citet{klypin_etal01} for galaxy-size halos and several
recent studies
\citep{power_etal03,ascasibar_etal03,fukushige_etal03,hoeft_etal03,hayashi_etal03}.
It is still not clear whether the density profiles in our simulations
are consistent with density distribution of observed clusters. We
note, however, that at the scales probed in observations the slope is
not expected to be shallower than $-1$.

The asymptotic value of the slope has been a subject of much numerical
effort in the last several years. Our results indicate that a
universal asymptotic slope may not exist. We should note that the
resolution of current dissipationless simulations is sufficiently high
to converge on the density profile at scales smaller than the size of
a typical central galaxy in clusters and groups ($\sim 30-50$~kpc).
Further improvement in profile modeling should therefore include
realistic dynamics and cooling of the baryonic component as
contraction of gas is expected to significantly modify dark matter
distribution at these scales.

One of the most interesting results of our study is existence of
systems with density profiles that can be well described by a power
law $\rho\propto r^{-\gamma}$ with $\gamma$ ranging from $\approx
-1.5$ to $\approx -2$. All of these systems are still in their rapid
mass growth stage and experienced a recent major or minor merger.
Remarkably, these halos maintain the power law density profiles at
earlier epochs out to at least $z\sim 1.5$.  The relatively shallow $\gamma>-2$
power law slopes result in low concentrations as the scale radius
where the density profiles reaches the slope of $-2$ is at large
radii.  There are also indications that the profiles of all clusters
are power law like during their rapid mass growth stages. We find, for
example, that the power law provides an increasingly better fit with
increasing redshifts for all of our clusters. At early epochs
($z\gtrsim 1.5-2$), the power law fit is always either comparable or
better than the NFW, M, and JS analytic profiles. We did not find any
correlation of the power slope with the details of the cluster MAH.
It would be interesting to look for such correlations using a larger
sample of objects.

When the mass growth slows down at $z>z_{\rm f}$, an
outer steeper density profile is built up. As pointed out by
\citet{zhao_etal03a}, the difference in density profiles during the two
mass accretion regimes may be due to more violent and thorough
relaxation during the period of rapid mass growth. Although
\citet{zhao_etal03a} focused on halo concentrations and did not
consider density profile shapes, their results are consistent with our
conclusions. In particular, they find that during the rapid mass
growth stage the circular velocity is nearly constant from the scale
radius to the virial radius. This behavior is consistent with a
power law density distribution with a slope close to $-2$.

In a recent study, \citet{ascasibar_etal03} find that objects which
experienced a recent merger event\footnote{Note that
  \citet{ascasibar_etal03} identify a recent merger by the presence of
  massive substructures within virial radii of their systems. This is
  different from our definition, which identifies major mergers
  directly from mass accretion tracks.} have lower concentrations and
steeper inner profiles than more relaxed systems. This is consistent
with our findings described above.  At the same time,
\citet{ascasibar_etal03} find that relaxed systems are better fit by
the NFW, while systems with a recent major or minor merger by the M
profile \citep[see also][]{ascasibar_03}. They thus associate a
particular shape of the profile with a recent merger history. In
contrast, our results show the shape of density profiles is set early
in the halo evolution and is usually stable over the past ten billion
years.  Clusters with density profiles best described by the NFW
rather than a JS at $z=0$, tend to have NFW-like profiles at earlier
epochs as well.  The reverse is also true. We tested this conclusion
using the high-resolution re-simulation of one of the clusters in our
sample.  Also, we do not find any correlation between the redshift of
last major merger (in our definition) or the formation redshift 
and the best fit analytic profile.

The origin of the distinctive density profile 
shape of the CDM halos remains poorly understood. Our results and
results of other recent studies indicate that the shape is tightly 
linked to the halo mass accretion history. During the period of 
rapid mass accretion the violent relaxation is significant and results
in a power-law like density distribution. This stage of evolution usually
occurs early when the universe is dense and builds up the inner dense
regions of halo. At this point, the logarithmic slope of the 
density distribution is shallower than $-2$ over a large fraction of the
halo volume and its concentration is small. 
At later epochs, as the mass accretion rate slows down, 
the outer regions of the halo are built while its central regions remains
nearly intact. This can be seen in Figure~\ref{fig:bestfit_CL7and5}
and Figures~10-13 of \citet{fukushige_makino01}. 
This picture can explain why the best fit analytic profiles tend to be
the same at various redshifts. The fits are sensitive to the density
distribution at small and intermediate ($\sim r_{\rm s}$) 
radii which are set early. However, it is still unclear which process(es)
determines a particular shape of the profiles. The key to understanding 
these processes appears to be in the details of early evolutionary 
stages of CDM halos, which will be the subject of a future study.

\acknowledgements We would like to thank Tony Tyson for stimulating
discussions which motivated this study, and Risa Wechsler for a careful reading of the draft
and many useful comments and suggestions.  This work was supported by
the National Science Foundation (NSF) and NASA under grants No.
AST-0239759 (CAREER) and NAG5-13274, by the NSF Center for
Cosmological Physics (CfCP) at the University of Chicago, and by the
NSF grant AST-0206216 at NMSU. We would like to thank NSF/DAAD for
supporting our collaboration. The simulations were performed at the
Leibniz Rechenzentrum Munich and the John von Neumann Institute for
Computing J\"ulich.

\bibliography{ms.bbl}

\end{document}